\documentclass[10pt,conference]{IEEEtran}

\usepackage{svg}
\usepackage{xcolor,colortbl}
\usepackage{float}
\usepackage{graphicx}
\usepackage{subfigure}
\usepackage{multirow}
\usepackage{xcolor}
\usepackage{colortbl}
\usepackage{tcolorbox}
\usepackage{fancybox}
\usepackage{listings}
\usepackage{hyperref}%参考文献和表格框
\hypersetup{colorlinks,linkcolor={blue},citecolor={blue},urlcolor={red}} %参考文献和表格框
\hypersetup{colorlinks=true, urlcolor=blue}
\usepackage{color}
\usepackage{diagbox}
\usepackage{soul}
\usepackage{algpseudocode}
\usepackage{hyperref}
\usepackage{framed}
\usepackage{amsmath}
\usepackage{pifont}
\usepackage{enumitem}
\usepackage{balance}
\usepackage[normalem]{ulem}
\usepackage{url}
\usepackage[T1]{fontenc}
\usepackage[utf8]{inputenc}
\usepackage[font=small,labelfont=bf,tableposition=top]{caption}
\usepackage{booktabs}
\usepackage{threeparttable}
\usepackage{tcolorbox}
\usepackage{fontawesome5}
\usepackage[ruled,vlined,linesnumbered]{algorithm2e}

\definecolor{grey}{rgb}{0.9,0.9,0.9}
\definecolor{lightgreen}{HTML}{bae4b3}
\definecolor{lightgrey}{HTML}{f0f0f0}
\definecolor{mygreen}{HTML}{31a354}
\definecolor{mygray}{HTML}{666666}

\makeatletter
\newcommand*\titleheader[1]{\gdef\@titleheader{#1}}
\AtBeginDocument{%
  \let\st@red@title\@title
  \def\@title{%
    \bgroup\normalfont\large\centering\@titleheader\par\egroup
    \vskip1.5em\st@red@title}
}
\makeatother
\newcommand{\xxx}{{\myfont Differential Prompting}\xspace}
\newcommand{\gptbaseline}{\textsc{BaseChatGPT}\xspace}
\newcommand{\pg}{{\myfont Program Generator}\xspace}
\newcommand{\ftg}{{\myfont Test Case Generator}\xspace}

\newcommand{\ft}{{\myfont FT-IA}\xspace}
\newcommand{\fc}{{\myfont FT-Ia}\xspace}
\newcommand{\ff}{{\myfont FT-ia}\xspace}
\newcommand{\pt}{{\myfont PT}\xspace}
\newcommand{\iat}{{\myfont IT}\xspace}

\newcommand{\ftest}{failure-inducing test case\xspace}
\newcommand{\ftests}{failure-inducing test cases\xspace}

\newcommand{\quixbugs}{{\mycode QuixBugs}\xspace}
\newcommand{\codeforces}{{\mycode Codeforces}\xspace}

\newcommand{\tszon}[1]{\mytodored{[Tsz On: #1]}}
\newcommand{\update}[1]{#1}

\newcommand{\scc}[1]{\mytodoblue{[scc: #1]}}

\newcommand{\mytodoblue}[1]{\textcolor{blue}{\ding{46}~{\sf}~#1}}
\newcommand{\mytodored}[1]{\textcolor{red}{\ding{46}~{\sf}~#1}}

\newcommand*{\mycode}{\fontfamily{lmtt}\selectfont}
\newcommand*{\myfont}{\fontfamily{LinuxBiolinumT-OsF}\selectfont}

\newcommand{\gptft}{28.8\%\xspace}
\newcommand{\xxxft}{75.0\%\xspace}
\newcommand{\pynguinft}{7.5\%\xspace}
\newcommand{\xxxvsgptft}{2.6X\xspace}
\newcommand{\xxxvspynguinft}{10.0X\xspace}

\newcommand{\xxxrv}{74.6\%\xspace}
\newcommand{\gptrv}{6.8\%\xspace}
\newcommand{\xxxrvgptrv}{11.0X\xspace}

\newcommand{\xxxcorrectfp}{5.0\%\xspace}

\newcommand{\correctintention}{91.0\%\xspace}
\newcommand{\numcorrectintention}{364\xspace}

\newcommand{\xxxgeneratecorrect}{74.6\%\xspace}

\newcommand{\chatgptnobugfound}{89.2\%\xspace}
\newcommand{\chatgptmoreinformationrequired}{6.8\%\xspace}

\newcommand{\xxxcodeforcesbuggyftimproved}{66.7\%\xspace}
\newcommand{\gptcodeforcesbuggyftimproved}{16.7\%\xspace}
\newcommand{\xxxcodeforcesbuggyftoriginal}{41.0\%\xspace}
\newcommand{\gptcodeforcesbuggyftoriginal}{7.0\%\xspace}
\newcommand{\xxxcodeforcesbuggyftvsbaselinesoriginal}{5.9X\xspace}
\newcommand{\xxxcodeforcesbuggyftvsbaselinesimproved}{4.0X\xspace}

\renewcommand{\baselinestretch}{.97} %修改行间距

\IEEEoverridecommandlockouts
\title{Nuances are the Key: Unlocking ChatGPT to Find Failure-Inducing Tests with Differential Prompting}
\titleheader{Accepted to the 38th IEEE/ACM International Conference on Automated Software Engineering}
\begin{document}

\DeclareRobustCommand*{\IEEEauthorrefmark}[1]{%
  \raisebox{0pt}[0pt][0pt]{\textsuperscript{\footnotesize #1}}%
}

\author{\IEEEauthorblockN{Tsz-On Li\IEEEauthorrefmark{a,b}, Wenxi Zong\IEEEauthorrefmark{c}, Yibo Wang\IEEEauthorrefmark{c}, Haoye Tian\IEEEauthorrefmark{d}, Ying Wang\IEEEauthorrefmark{c,a}\IEEEauthorrefmark{*}\thanks{\IEEEauthorrefmark{*}\small{Ying Wang and Shing-Chi Cheung are the corresponding authors.}}, Shing-Chi Cheung\IEEEauthorrefmark{a,b}\IEEEauthorrefmark{*}, Jeff Kramer\IEEEauthorrefmark{e}}
\IEEEauthorblockA{
  \textit{\IEEEauthorrefmark{a}The Hong Kong University of Science and Technology, Hong Kong, China}\\
  \textit{\IEEEauthorrefmark{b}Guangzhou HKUST Fok Ying Tung Research Institute, Guangzhou, China}\\
  \textit{\IEEEauthorrefmark{c}Northeastern University, Shenyang, China}\\
  \textit{\IEEEauthorrefmark{d}University of Luxembourg, Luxembourg, Luxembourg}\\
  \textit{\IEEEauthorrefmark{e}Imperial College London, London, United Kingdom}\\
  \{toli,scc\}@cse.ust.hk, iamwenxiz@163.com, yibowangcz@outlook.com, haoye.tian@uni.lu,\\ wangying@swc.neu.edu.cn, j.kramer@imperial.ac.uk
}}
	\maketitle
 \vspace{-7mm}

\begin{abstract}
% \TszOn{TODO}
% \scc{TODO}
% \ying{TODO}

%\scc{I wonder if we are using the right TSE template.} 

Automated detection of software failures is an important but challenging software engineering task. It involves finding in a vast search space the \ftests that contain an input triggering the software fault and an oracle asserting the incorrect execution. We are motivated to study how far this outstanding challenge can be solved by recent advances in large language models (LLMs) such as ChatGPT. 
% Automatically detecting software failure 
%to various software engineering tasks, such as bug diagnosis and program repair. 
%However, automatically finding a failure-inducing tes is known as a challenge. 
%With the emergence of a powerful large language model called ChatGPT, we took the first attempt to validate its effectiveness in addressing such a challenge.
%Recent emergence of a powerful large language model called ChatGPT has shed light on overcoming this challenge. 
%We take the first step to evaluate ChatGPT's effectivenss in identifying a failure-inducing test. 
However, our study reveals that ChatGPT has a relatively low success rate (\gptft) in finding correct failure-inducing test cases for buggy programs. %It may also incorrectly suggest failure-inducing tests for many correct programs. \tszon{For instance, in our study, ChatGPT incorrectly suggested a failure-inducing tes for every correct program}
%We also observe that it incorrectly suggests a failure-inducing tes for all the correct programs.
%\ying{We also observed that it incorrectly suggested a failure-inducing tes for all the correct programs.}. %a given buggy program, we observed that ChatGPT is weak at finding its failure-inducing tes (e.g., only \gptft failure-inducing tests found by ChatGPT are true positive, the remaining failure-inducing tests mostly raised false positive). 
%We discovered it is because 
A possible conjecture is that finding failure-inducing test cases requires analyzing the subtle differences (nuances) between the tokens \update{of} a program's correct version and those for its buggy version. %two similar sets of tokens that capture the correct and the buggy versions of a program. 
When these two versions have similar sets of tokens and attentions, ChatGPT is weak in distinguishing their differences. %The task may not readily fit the design of ChatGPT, which infers upcoming tokens (e.g., a failure-inducing tes) from previous tokens (e.g., a program). %The main reason is that ChatGPT is limited in identifying semantic nuances (i.e., subtle semantic differences between similar input tokens), while a bug is essentially a semantic nuance between a buggy program and a bug-free program. 
%Hence, ChatGPT is inevitably weak at identifying bugs and thus corresponding failure-inducing tess.
% Such an insensitivity causes ChatGPT unable to detect buggy syntax from a program, and thus unable to generate a test case to make a bug manifest.

We find that ChatGPT can successfully generate failure-inducing test cases when it is guided to focus on the nuances. Our solution is inspired by an interesting observation that ChatGPT could infer the intended functionality of buggy code if it is similar to the correct version. Driven by the inspiration, we develop a novel technique, called \xxx, to effectively find failure-inducing test cases with the help of the compilable code synthesized by the inferred intention. Prompts are constructed based on the nuances between the given version and the synthesized code.
We evaluate \xxx on {\mycode QuixBugs} (a popular benchmark of buggy programs) and recent programs published at {\mycode Codeforces} (a popular programming contest portal, which is also an official benchmark of ChatGPT). We compare \xxx with two baselines constructed using conventional ChatGPT prompting and \textsc{Pynguin} (the state-of-the-art unit test generation tool for Python programs). 
Our evaluation results show that for programs of \quixbugs, \xxx can achieve a success rate of \xxxft in finding failure-inducing test cases, outperforming the best baseline by \xxxvsgptft. For programs of \codeforces, \xxx's success rate is \xxxcodeforcesbuggyftimproved, outperforming the best baseline by \xxxcodeforcesbuggyftvsbaselinesimproved.
%Our evaluation result shows that \xxx is much more effective in finding a failure-inducing tes than the baselines. For instance, \xxx's success rate of finding a failure-inducing tes is \xxxft, at least \xxxvsgptft more than all the baselines.

% Our insight is that ChatGPT's insensitivity to semantic nuances is a double-edged sword: although it is unfavourable to bug detection, it is favourable to another closely-related task \textit{code intention inference}. 
% \let\thefootnote\relax\footnote{{There is no number in this footnote}}
\end{abstract}


\section{Introduction}\label{sec:intro}
 \begin{quote}
     \emph{\textcolor{gray}{\faIcon
 {quote-left}} \textbf{Testing leads to failure, and failure leads to understanding.}} \textcolor{gray}{\faIcon
 {quote-right}} \hfill - Burt Rutan
    %  \begin{flushright}
    %  - Burt Rutan
    %  \end{flushright}
 \end{quote}

\emph{\textbf{Problem and Challenges.}}
Finding \ftests is a main objective in software testing. %A test suite typically consists of two types of test cases: passing test cases and \ftests.
%In practice, passing test cases heavily outnumber \ftests regarding a fault, which hampers the effectiveness of program debugging.
%Bugs in software can cause huge financial loss~\cite{wong2017more}. 
%To detect, diagnose and fix bugs, a developer often needs to find \ftests (i.e., test cases that trigger a program failure). 
However, finding such test cases is challenging in practice. 
First, it is because a developer needs to search for a test input that can trigger a program failure, while the search space of such a test input is vast~\cite{fraser2015does}. Second, the developer needs to construct a test oracle to automatically detect a program failure, while constructing a test oracle is often an undecidable problem~\cite{ibrahimzada2022perfect}.

% the search space of program inputs is typically large, and constructing a test oracle for detecting a program's failures is often an undecidable problem.

% Since ChatGPT is powerful, 
%Recently, a powerful Large Language Model (LLM) called ChatGPT~\cite{chatgpt} has emerged. 
Recent studies~\cite{sobania2023analysis,xia2023conversational} have reported the potential of using large language models (LLMs) like ChatGPT~\cite{chatgpt} for software engineering tasks, such as automated program repair. In this paper, we conduct the first study to investigate the use of ChatGPT in finding \ftests for a given program. We do not focus on finding test cases that achieve high code coverage because it is not necessarily correlated with fault detection abilities. In addition, finding test cases whose assertions are incorrectly generated is not useful even if they achieve high code coverage. Our study aims to find test cases with failure-inducing test inputs and correctly specified assertions. However, using ChatGPT to find such test cases is challenging.

We experiment using ChatGPT in a direct manner with both buggy and correct Python programs in {\mycode QuixBugs}~\cite{lin2017quixbugs}, which is a common benchmark used by recent works~\cite{xia2023conversational,sobania2023analysis} to study the use of LLMs for SE tasks. We prompt ChatGPT in two steps, following a recently proposed prompting convention~\cite{sobania2023analysis}. For each program, we first ask ChatGPT to determine whether a given program contains bugs. When it replies yes, we ask it to generate a \ftest. Our experiment yields two findings: first, ChatGPT has a low success rate (\gptft) in finding \ftests for buggy programs; %\scc{How do we prompt ChatGPT?} %Worse, we conducted another experiment by requesting ChatGPT to find \ftests for each correct program of \textsc{Quixbugs}. 
second, ChatGPT often considers correct programs buggy and suggests \ftests. 
%\scc{FYI, the explanation of direct prompting can be made in later sections in order to reduce the burden in understanding the main points in Introduction. I think it serves the purpose by using ``ChatGPT'' instead of ``direct prompting of ChatGPT'' in this paragraph because the meaning should be clear even without mentioning direct prompting. In addition, it can be confusing to say that direct prompting follows an existing work and at the same time we argue that we carry out the first study on finding \ftests. Please note how I have rephrased it. Please comment out this note when you have read it.} %\scc{How do we prompt ChatGPT?} %The result shows that ChatGPT has a high false positive rate. Overall, these two experiments indicate that ChatGPT is far from practical for finding \ftests.

% The remaining \gptff of the generated test are false positive. These false positive tests essentially hinder developers from detecting, diagnosing and fixing actual bugs. \tszon{using precision and recall instead}.

%In this paper, we study the ways to overcome the weaknesses of ChatGPT and propose the \xxx methodology. 
A possible explanation of the findings is that ChatGPT is essentially a Transformer (a Deep Learning model architecture commonly adopted by LLM)~\cite{chatgpt}, which is found to be insensitive to nuances (i.e., subtle differences between two similar sequences of tokens)~\cite{zhang2020semantics,zhang2021fast,jiang2021lightxml,chang2020taming}. In other words, when a sequence of input tokens varies slightly, LLMs often make similar inferences. This property allows LLMs to tolerate some noise in input tokens.
% Such a property is usually favorable to many NLP tasks, where variation of individual tokens (e.g., auxiliary words) should not affect an inference result (e.g., \tszon{need an example in NLP}). 
Nevertheless, a bug is essentially a nuance between a buggy program and its fixed version~\cite{patra2021semantic,tan2014bug,liu2019avatar}. 
Hence, an LLM (including ChatGPT) cannot reliably detect the existence of bugs in a program and is therefore weak in finding correct \ftests.

\begin{figure}[t!]
	\centering
		\vspace{-2mm}
	\includegraphics[width=0.5\textwidth]{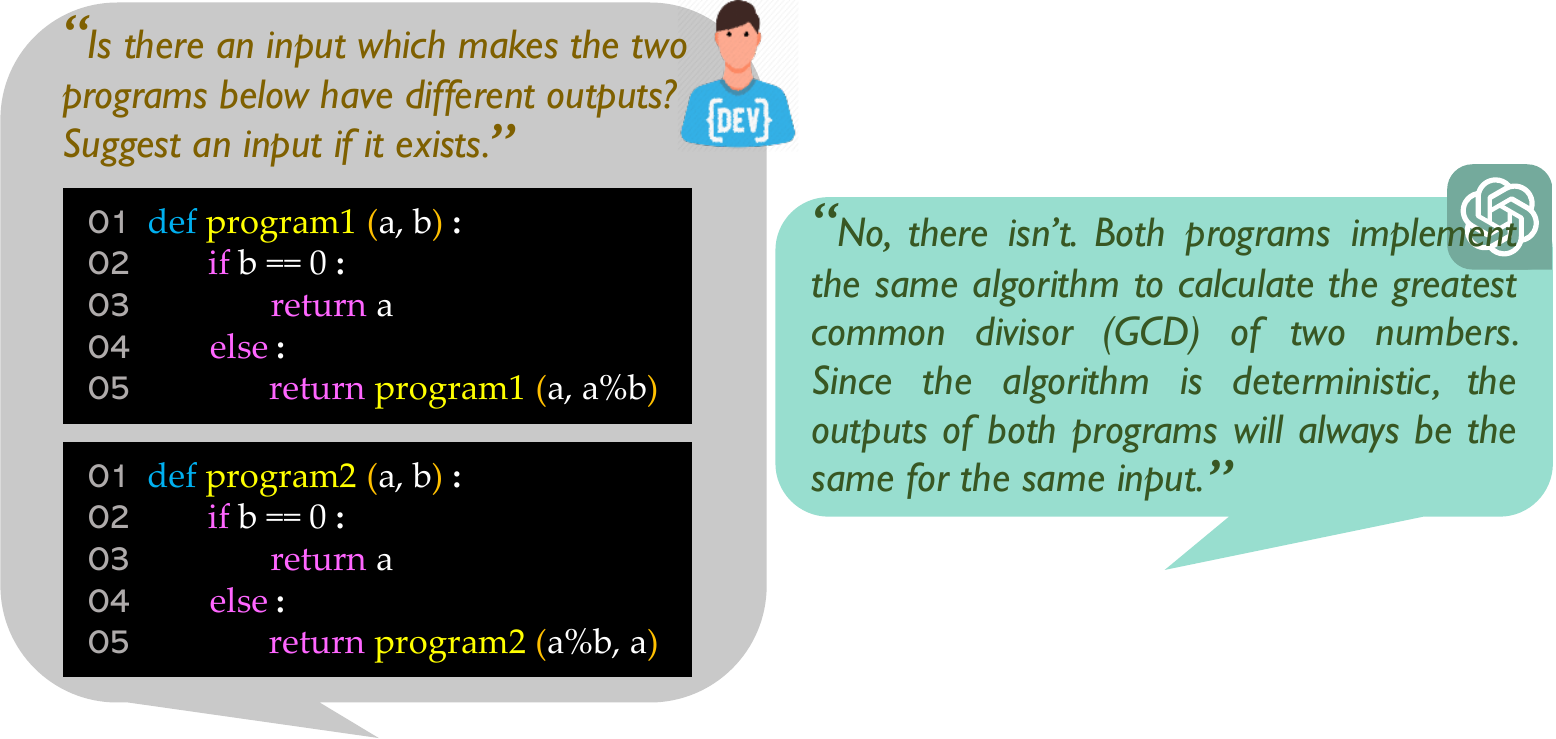}
	\vspace{-6mm}
 \caption{An illustrative example for ChatGPT's weakness.%, where \emph{program1} and \emph{program2} are non-buggy and buggy implementations of the greatest common divisor, respectively.
 }
	%\caption{ChatGPT mistakenly considered that program1 (buggy version) and program2 (bug-fixed version) always have the same output for the same input.}
	\label{fig:listing1}
	\vspace{-6mm}
\end{figure}

Figure~\ref{fig:listing1} shows an illustrative example that consists of two buggy implementations of greatest common divisor. We ask ChatGPT to generate an input that can make the two programs have different outputs. When $a$=12 and $b$=20, the outputs of the two programs differ. However, ChatGPT responds that no input can make the two programs have different outputs. ChatGPT shows its insensitivity to the presence of nuances in this example.

\emph{\textbf{Insights.}} To tackle the weakness of ChatGPT in finding \ftests, we leverage two insights.
%To address ChatGPT's limitation, we have two major insights. 
First, the task of finding a \ftest for a program can be much facilitated if its intention is known. %is closely related to the task of inferring a program's intention. %(a.k.a., program intention inference).
It enables us to focus our search for \ftests on those that exhibit behaviors different from the program intention.
%It is because a \ftest essentially detects a program's behavior that deviates from the program's intention. 
%Hence, by getting a program's intention and comparing it against the program's implemented behaviors, the program's failure and the corresponding \ftest can be much easier to find. 

Second, the weakness of ChatGPT at recognizing nuances can facilitate ChatGPT to infer a program's intention, because such a weakness allows ChatGPT to ignore the presence of a bug (i.e., nuances) when inferring the program's intention. To validate this insight, we conduct an experiment on {\mycode QuixBugs} by requesting ChatGPT to infer a buggy program's actual intention.
% (i.e., the intention of the program without the bug).
Our experiment result shows that ChatGPT correctly infers the actual intention of \correctintention buggy programs. 

\emph{\textbf{Approach.}} Leveraging these two insights, we propose \xxx, a new paradigm for finding \ftests. \xxx divides the task of finding \ftests into three sub-tasks: program intention inference, program generation, and differential testing. 
Specifically, given a program-under-test (PUT), \xxx first asks ChatGPT to infer the intention of a PUT. ChatGPT is then prompted to generate multiple compilable programs that have the same intention as the PUT. Essentially, these programs are alternative implementations of the PUT. %Although these programs may not necessarily be correct, they are unlikely to suffer from the same PUT's bug. 
Since these programs may also be buggy, we consider only those test cases that deliver consistent results across these programs. Here, we assume that these generated programs are unlikely to commonly suffer from the same bug. Therefore, if one of these test cases induces PUT to deliver a different result, the test case will likely be failure-inducing. In other words, we apply differential testing between PUT and the programs generated by ChatGPT based on the inferred program intention. %, because they are generated based on the intention of the PUT, instead of the actual implementation of the PUT. 
%In other words, given the same input, any inconsistency between the outputs of the PUT and the generated programs indicates the PUT's failure. Hence, \xxx performs differential testing with the PUT and the generated programs to find \ftests. 
%Finally, \xxx conducts differential testing for PUT with the generated programs, in order to find a failing test for PUT.
By doing so, \xxx bypasses the weakness of ChatGPT at recognizing nuances, while leveraging its capability of inferring program intention. Note that \xxx may find incorrect \ftests in the few scenarios when programs generated by ChatGPT commonly suffer from the same bug. We will evaluate this and suggest measures to mitigate this in our evaluation. 
% Hence, PUT will behavior differently with these alternative implementations. Such a difference in behaviors enable differential testing, which generate tests to detect the difference in behaviors between a program-under-test and its alternative implementation. The test which triggers such a difference would be a failing test for program-under-test.

\emph{\textbf{Evaluation.}} We evaluate \xxx on {\mycode QuixBugs} and \codeforces~\cite{codeforces}. %(a popular programming contest portal and an official benchmark of ChatGPT). 
\codeforces publicly releases all buggy or correct programs submitted by contest participants. %, and allows these programs to be accessed by the public. 
%These programs essentially form a benchmark similar to \quixbugs. 
We select seven programs that were submitted after September 2021 (the cut-off date of ChatGPT's training dataset) for evaluation, in order to mitigate the threat of data leakage.
We compare \xxx against two baselines: ChatGPT and Pynguin~\cite{lukasczyk2022pynguin} (the state-of-the-art unit test generation tool for Python program). Our evaluation results show that for buggy programs of \quixbugs, the success rate of \xxx in finding \ftests is \xxxft, outperforming the best baselines (\gptft) by \xxxvsgptft. For correct programs of \quixbugs, \xxx only incorrectly suggests a \ftest for \xxxcorrectfp of the programs. In comparison, the two baselines incorrectly suggest a \ftest for every correct program. %\scc{The last sentence is confusing. We need to explain the meaning of direct use. For example, how do we prompt the ChatGPT? Reviewers may wonder if it is a natural result if the prompt asks ChatGPT to generate failure-inducing test cases for a correct version. This paragraph is important but confusing. It seems that we evaluate only on Quixbugs, and compare with baselines only on this dataset. We have not carried out any experiments to address the data leakage issue. Similar issues occurred at Contributions. If we plan to include Codeforces results to be filled later, there should be some placeholders in the writing. I stop reviewing here because I feel that important content has not been added. Please let me know when Introduction is ready for my review. In addition, the detailed justification for selecting Quixbugs should be left to the experiment design. I am surprised that Codeforces is not mentioned anywhere in Introduction and Abstract.} 
For buggy programs of \codeforces that have similar complexity with buggy programs of \quixbugs, \xxx’s success rate is \xxxcodeforcesbuggyftimproved, outperforming the best baselines (\gptcodeforcesbuggyftimproved) by \xxxcodeforcesbuggyftvsbaselinesimproved. For the correct programs, \xxx only incorrectly suggests \ftests for 1 out of 7 programs, while the baselines incorrectly suggest \ftests for all the seven programs.
% Besides, 
% Besides, none of the failing tests found by \xxx is false positive. In comparsion, on average \gptff and \pynguinff of failing test found by the baselines (ChatGPT and Pyguin respectively) are false positive. For correct programs, \xxx's false positive rate is merely \xxxfffree, much less than the baselines at least \fffreediff X more than the baselines.
% The first baseline is directly requesting ChatGPT to generate a failing test (ChatGPT-Direct). The second baselines is Pynguin, a well-known automatic unit test generation tool. 
% For each baselines, we generate 
% Our evaluation results shows that by generating the same number of test cases, 

% Besides, we applied \xxx and ChatGPT to each of the forty bug-free programs in Quixbugs. Our evaluation results shows that on average only \xxxff-free failing tests identified by \xxx

% \todo{x}\% more test cases which trigger true failure than the baselines, and \todo{x}\% less failing test case which trigger true false than the baselines. \todo{Some suprising and significant results here}.

\emph{\textbf{Contributions.}} To summarize, this paper makes the following three contributions:
\begin{itemize} [leftmargin=*, topsep=1pt, itemsep=1pt]
  \item \emph{\textbf{Originality:} To the best of our knowledge, we conducted the first study to investigate ChatGPT's effectiveness in finding \ftests.} Our findings help understand ChatGPT's \update{limitations} in doing so, and inspire future studies in this area. %shows that ChatGPT is limited to do so. We pointed out a possible reason is that ChatGPT's model architecture makes ChatGPT insensitive to nuances. 
  % Our findings not only points out ChatGPT's limitation in finding, but can also 
  
  \item \emph{\textbf{Technique:} We propose \xxx, a new paradigm for finding \ftests.} \xxx unlocks ChatGPT's power in detecting software failure, by formulating the task of finding \ftests into program intention inference, program generation, and differential testing. By doing so, \xxx turns ChatGPT's weakness into a strength in finding \ftests.
  The evaluation result  shows that \xxx 
  notably outperforms the state-of-the-art baselines on \quixbugs and \codeforces. 

  \item \emph{\textbf{Evaluation:} We implement \xxx and evaluate it  on \quixbugs and \codeforces.} Our evaluation result shows that \xxx's success rate outperforms the best baseline by \xxxvsgptft on \quixbugs and \xxxcodeforcesbuggyftvsbaselinesimproved on \codeforces. We make our research artifact and experimental data available.
    %\scc{Implementation, evaluation and making the research artifact available are considered contributions.}
  
  % , a new paradigm for finding \ftests.} \xxx unlocks ChatGPT's power in detecting software failure, by formulating the task of finding \ftests into program intention inference, program generation, and differential testing. By doing so, \xxx turns ChatGPT's weakness into a strength in finding \ftests.
  % The evaluation result shows that \xxx 
  % notably outperforms the state-of-the-art baselines on \quixbugs and \codeforces. 

  % \item \emph{\textbf{Data Availability.}}  \hl{We provide a reproduction package at \textbf{\emph{http://}} to facilitate future research. The package includes (1) a dataset ... ; (2) ....; (3)  .} 

  %\ying{Data Availability should be explicitly described in our paper. It is better to put it at the end of Intro.} \scc{Agree with Ying. Or Put it in a separate Section after the Conclusion, following the ESEC/FSE recommendation. Data Availability is NOT an item under Contributions.}
  
 % \item \emph{\textbf{Evaluation:}} The evaluation result shows that \xxx 
 % notably outperforms the state-of-the-art baselines on \quixbugs and \codeforces. 
%  \xxx's source code and experiment data is available at \emph{\textbf{http://}} \todo{url}. 
%\scc{Do we intend to release our experiment datasets to facilitate related future studies? If so, why don't we include it as a contribution? Please refer to existing papers how to claim contributions.} 
%\ying{We should give the url before submitting.}
\end{itemize}

\emph{\textbf{Promising Application Scenarios.}} We acknowledge that \xxx in this paper works best in finding \ftests for simple programs (i.e., programs that have less than 100 lines of code). However, the improvement that it offers over state-of-the-art approaches is an essential step towards finding \ftests for larger software. Larger software is inevitably associated with more sophisticated program intention/requirement, which can be difficult to describe precisely in natural language. In addition, the input search space can be much larger. As a result, \ftests are more difficult to find.
  
  % We also discuss how \xxx can be extended to large programs in the discussion .

  Besides, \xxx's effectiveness in finding \ftests also benefits education for software engineering. Specifically, our evaluation result implies that \ftests found by \xxx can accurately detect failures of programming beginners' programs (which are usually simple programs). Previous studies~\cite{lahtinen2005study,kelleher2005lowering, perretta2022use} on software engineering education suggest that those \ftests can facilitate a programming beginner's understanding of programming concepts, and help the beginner effectively learn debugging practices. \update{Hence, like many software engineering techniques proposed for programming assignments~\cite{yi2017feasibility,bhatia2018neuro,hu2019re}, \xxx can be adopted by beginners for learning programming.}

\begin{figure*}[t!]
	\centering
	%	\vspace{-3mm}
	\includegraphics[width=1.05\textwidth]{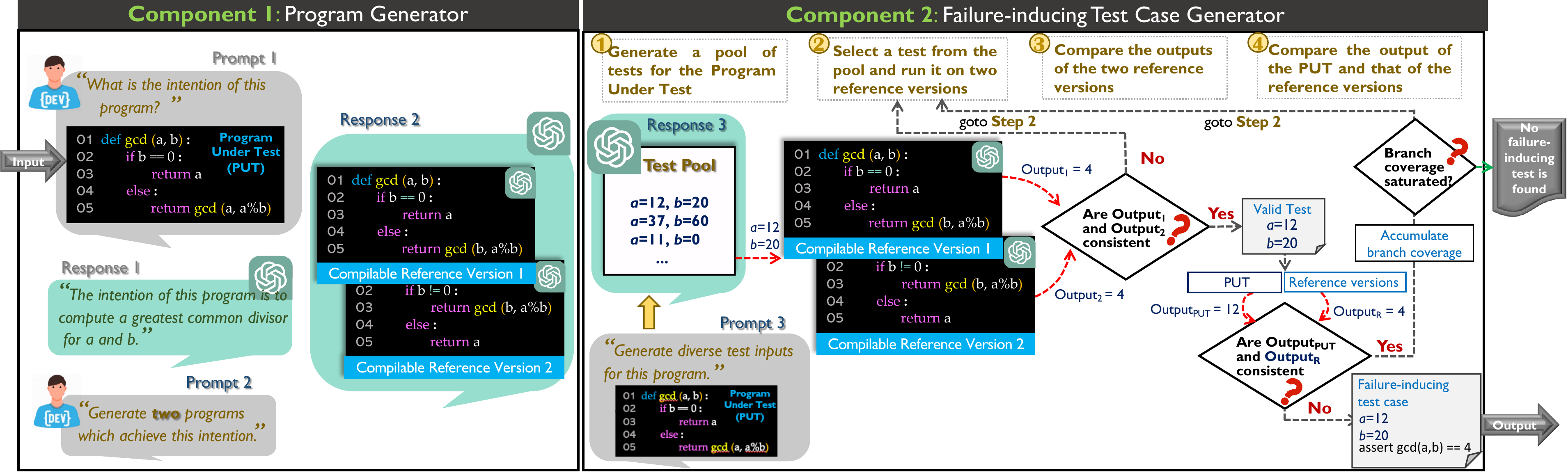}
	\vspace{-5mm}
	\caption{Workflow of \xxx} %\ying{Please check if it is ok.}
	\label{fig:workflow}
	\vspace{-7mm}
\end{figure*}

% Figure~\ref{fig:Fig3} shows an overview of our approach, including the following three stages:
\section{Preliminaries}\label{sec:formulation}
%\ying{Dear professor, Section 2 is ready for your edit.} \scc{Done.}
% Our prompting convention follows a previous work. First is there, 

We use a running example in Figure~\ref{fig:listing1} to illustrate the task of finding \ftests. %We first describe the running example. 
%{\mycode program1} is a bug-free implementation of greatest common divisor, while {\mycode program2} is a buggy implementation of greatest common divisor. 
In the example, {\mycode program1} and {\mycode program2} differ in their last return statement, which is {\mycode return program1(a,a\%b)} for the former and {\mycode return program2(a\%b,a)} for the latter. Essentially, they swap the order of arguments.   %of {\mycode program2}, where the order of {\mycode a\%b} and {\mycode b} have been swapped). 
This makes the output of {\mycode program1} and {\mycode program2} different for certain inputs. For instance, when $a$ = 12 and $b$ = 20, {\mycode program1} outputs 12 while {\mycode program2} incorrectly outputs 0. However, {\mycode program1} intends to implement a greatest common divisor (GCD) program. Hence, given $a$ = 12 and $b$ = 20, the output of {\mycode program1} should be 4.

% \subsection{Finding a Failing Fest}

A test case found by \xxx, ChatGPT or Pynguin~\cite{lukasczyk2022pynguin} can fall into one of the five categories. % (described in \S\ref{sec:intro}). 
%Specifically, the evaluation metric is \textit{success rate}, the number of correct \ftests found by \xxx (or a baseline) divided by the total number of attempts made (each attempt is supposed to find one failing test, see \S\ref{sec:eva:rq1:setup} for more details).
% of \xxx and the baselines: precision and recall.
% Only one of the category is considered a true positive in the problem of finding a failing test. The remaining four are considered false positives.

%A \textbf{\emph{failure-inducing test case}} can fall into three categories:
\begin{itemize} [leftmargin=*, topsep=1pt, itemsep=1pt]
  \item \textbf{\textit{Correct failure-inducing test case} (\emph{\ft})}. %(``true failure'' is a program failure induced by fault(s) residing in a program).
  An \emph{\ft} consists of (1) a failure-revealing test input \textit{I} that triggers PUT to give an incorrect output, and (2) a correct assertion \textit{A} that identifies the incorrect output. So, both failure-inducing input and assertion are correctly generated. %statement that asserts the PUT's output against the correct output of the PUT. 
  An example is a test case that executes {\mycode program1(12,20)} and asserts its expected output to be 4. It correctly reveals a bug in {\mycode program1}. A failing test triggered by an \emph{\ft} is considered a \textit{true failure}. 
 
  %For ease of writing, we propose an acronym \ft to represent a test case of this category. 
 %Specifically, ``{\mycode FT}'' represents \ftest; other categories of test cases such as passing test cases and invalid test cases are represented by ``{\mycode PT}'' and ``{\mycode IT}'' respectively (these categories of test cases are discussed in following paragraphs). 
 %Besides, ``{\mycode I}'' means a test input is fault-inducing, while ``{\mycode i}'' means a test input is not fault-inducing. 
 %Finally, ``{\mycode A}'' means an assertion statement captures a PUT's correct expected output, while ``{\mycode a}'' means the otherwise. 
 %We follow this convention to derive an acronym for other categories of test cases.

  \item \textbf{\textit{Coincidental failure-inducing test case} (\emph{\fc})} %(``Coincidental failure'' is a test failure that coincidentally occurs with a true failure).
  An \emph{\fc} is similar to an \emph{\ft} except its assertion \textit{a} incorrectly specifies the expected output. A failure is coincidentally observed because the incorrect output given by the PUT happens to be different from the incorrectly specified expected output in the assertion. In other words, the failure-inducing input is correctly generated but the assertion is not. %consists of (1) a failure-revealing test input (I) that triggers PUT t a PUT's failure, and an assertion statement that asserts the PUT's output against an incorrect output. 
  An example is a test case that executes {\mycode program1(12,20)} but incorrectly asserts its expected output to be a value (e.g., 2) other than 4 (correct output) and 12 (incorrect output of PUT). A failing test triggered by an \emph{\fc} is considered a \textit{coincidental failure}. %{\mycode program2}'s output to be any value other than 1 (correct output) and 37 (actual output). In this case, \fc reports a test failure because {\mycode program2}'s actual output is not equal to the asserted output. 
%However, since \fc's assertion statement is incorrect, 
An \emph{\fc} does not help fault diagnosis and program repair. It rejects correct patches. %inevitably hinders a developer from diagnosing or repairing a fault. 
%Hence, \fc is considered an \textit{incorrect \ftest}.

 \item \textbf{\textit{False failure-inducing test case} (\emph{\ff})} %(``False failure'' is a test failure that occurs when true failure does not occur, so false failure is essentially a false alarm). 
An \emph{\ff} reports a false alarm. The test case consists of (1) a non-failure-inducing test input \textit{i}, and (2) an assertion \textit{a} that incorrectly asserts the expected output. So, both the failure-inducing input and the assertion are incorrectly generated. %PUT's output against an incorrect output. 
%suppose when $a$ = 17 and $b$ = 0, {\mycode program2} correctly returns 17. 
An example is a test case that executes {\mycode program1(17,0)} and incorrectly asserts its expected output to be 18. A failing test triggered by an \emph{\ff} is considered a \textit{false failure}. %, so \ff is essentially an incorrect \ftest.

\item \textbf{\textit{Passing test} (\emph{\pt})}. A \emph{\pt} results in a passing test. %, in which the actual outputs match the expected outputs.%, so it is considered an incorrect \ftest.

  \item \textbf{\textit{Illegal argument test case} (\emph{\iat})}. An \emph{\iat} consists of an illegal test input violating the designated argument type. %\iat is also considered an incorrect \ftest.

\end{itemize}

% A \ft is considered correct \ftest.
%Hence, \fc is considered an \textit{incorrect \ftest}.
%so \ff is essentially an incorrect \ftest.
Note that a test case with a correct assertion (A) and a non-failure-inducing test input (i) cannot trigger a failure. Such a test case is either PT or IT. Out of the five categories, an \emph{\ft} is considered a \textbf{\textit{correct \ftest}}, while test cases in the other categories are considered \textbf{\textit{incorrect \ftests}}. 
The correctness criterion allows us to design an evaluation metric (e.g., \emph{success rate}) to compare the effectiveness of \xxx with the baselines (see \S\ref{sec:eva:rq1:setup}). 

\section{Methodology}\label{sec:methdology}
\xxx is designed to correctly find \ftests for a PUT. Figure~\ref{fig:workflow} shows the workflow of \xxx: it accepts a PUT and outputs either a \ftest or a message that it cannot find a \ftest. 
\xxx consists of two main components: \pg (\S\ref{sec:pg}) and \ftg (\S\ref{sec:ftg}).

% Overall speaking, \pg's objective is to generate multiple alternative implementations (reference versions) of the PUT, and \ftg leverage these reference versions to perform differential testing for PUT , in order to identify a failing test (\S\ref{sec:ftg}). \pg and \ftg synergistically addresses two 

% \textbf{Step1: Equivalence Modulo Input (EMI) Mutator} takes a code snippet as an input, mutating the syntax of the code snippet, and output a mutated code snippet. Specifically, EMI~\cite{sun2016finding} is a code mutation technique which generates a new code whose syntax is different from the original code, while the behavior (i.e., input and output mapping) is identical to the original code. EMI has been widely adopted in compiler testing, and we take the first step is apply EMI to program repair, thanks to the conversational property of CHatGPT. \tszon{Why no one has adopted EMI for program repair MR before? is there other program repair MR?}

% Following previous works~\cite{sun2016finding}, we chose three mutation strategy to mutate a code snippet's syntax: \textsc{Always False Conditional Block},  \textsc{Always False Conditional Block}which adds a conditional block to the code snippet, while the predicate of the conditional block can never be satisfied (e.g., if false). Hence, the conditional block will never be entered. To be continued.

% \step1's objective is generate multiple mutants for metamorphic testing (see \S\ref{sec:mr})

\subsection{\pg}\label{sec:pg}

\pg's objective is to generate multiple ``reference versions'' of a PUT, each of which provides an alternative implementation of the PUT. However, our experiment on {\mycode QuixBugs} programs finds that only 6.8\% of the programs generated by ChatGPT are correct when it is prompted to generate bug-free versions of the {\mycode QuixBugs} programs (see \S\ref{sec:rq2:finding}). Therefore, such a strawman approach to \update{generate} reference versions does not work well. In the following, we present a more effective mechanism for \pg. %: \textcolor[rgb]{0.41,0.41,0.41}{\emph{\dotuline{a reference version is a PUT's alternative implementation}}}. 
%Note that a reference version does not have to be totally bug-free. As long as a reference version has different failure(s) from a PUT, the reference version can still be leveraged to find \ftests via differential testing (to be further discussed in \S\ref{sec:ftg}).
%(e.g., the reference versions do not have the PUT's failure). 
%With the reference versions, \xxx can perform differential testing in order to find \ftests (see \S\ref{sec:ftg} for more details). 
% \pg takes PUT as an input, and outputs references of the PUT to \ftg. 
% \pg is divided into two steps : Code Intention Inference and Program Generation (discussed in \S\ref{sec:pgworkflow}).

\subsubsection{Overview of \pg}\label{sec:pg:overview}

As pointed out in \S\ref{sec:intro}, ChatGPT can be insensitive to code nuances. It is a double-edged sword. On the one hand, it hinders ChatGPT from identifying bugs arising from these nuances. On the other hand, it lets ChatGPT infer the intention of a PUT despite the presence of these bugs. 
% Our experiment result shows that ChatGPT correctly infers the actual intention of \correctintention buggy programs
%Such an intention can be leveraged to generate reference versions of a PUT.
Leveraging this insight, \pg divides the task of generating reference versions into two steps. It first leverages ChatGPT to infer the intention of a PUT. Then, it leverages ChatGPT to generate multiple compilable reference versions of the PUT based on the inferred intention. These reference versions will be used by \ftg for differential testing. 
%(see \S\ref{sec:pg:worflow} for detailed workflow). 
By doing so, \pg no longer relies on the generation of correct patched programs from buggy code, and bypasses the weakness using the strawman approach. 

We consider a reference version generated by ChatGPT from the inferred intention is \emph{good} if the reference version does not suffer from the same bug(s) as PUT. It is a necessary condition for \xxx to deduce the correct expected output of a \ftest using the reference version. Our experiment results show that \xxxgeneratecorrect of the reference versions generated by \pg using the inferred intention are good %(i.e., do not have a bug), outperforming the strawman approach by \xxxtogptgeneratecorrect  
(see \S\ref{sec:eva:rq3}). %Indeed, \xxx does not need to rely on generating correct reference versions. 
The relaxation of the correctness requirement for generated reference versions allows \xxx to successfully find \ftests for most program subjects in \quixbugs~\cite{lin2017quixbugs}, significantly outperforming the baselines %Although some reference versions generated by \pg contain bug(s), \xxx is still able to find \ftests for PUT with these reference versions using differential testing 
(see \S\ref{sec:eva:rq1}).
%, thus finding \ftests. 

% Our insight to tackle such a limitation is that ChatGPT's insensitivity to nuances is a double-edged sword: although such an insensitivity hinders ChatGPT from identifying a bug, it benefits ChatGPT from performing another closely related task: code intention inference (\S\ref{sec:intro}). Code intention inference essentially requires ChatGPT to be insensitive to the presence of bugs (i.e., nuances) in a PUT, so that ChatGPT can infer the actual intention of the PUT. 

%We conducted an experiment with {\mycode Quixbugs} to evaluate ChatGPT's effectiveness in program intention inference (details about the experiment setup can be found in \S\ref{sec:rq2}). 
%In addition, 

% , and \xxx's success rate in generating a correct reference version is 

% Although there is a tiny proportion (\incorrectintention) of programs which \pg cannot infer the intention correctly, the incorrect intention can be repaired with feedbacks provided by \xxx's user regarding the failing test returned by \xxx. The details about how \xxx repairs the inferred intention are discussed in \S\ref{sec:discussion}. 

\subsubsection{Illustration of \pg's workflow}\label{sec:pg:worflow}

% To 
% \pg is divided into two steps: Code Intention Inference and Program Generation.

\emph{Component 1} in Figure~\ref{fig:workflow} illustrates \pg's workflow. Suppose that the buggy {\mycode gcd(a,b)} function described in \S\ref{sec:formulation} is the PUT, 
%(the {def \mycode gcd(a,b)} function shown in \emph{Prompt 1}). 
% The {\mycode gcd(a,b)} function is actually buggy because the last line of the code should be {\mycode return gcd(a\%b,b)} instead of returning {\mycode gcd(b,a\%b)}. 
%Suppose the developer inputs this function to \xxx, 
\pg first requests ChatGPT to infer the intention (as shown in  \emph{Prompt 1}). After ChatGPT returns an inferred intention (\emph{Response 1}), \pg requests ChatGPT to generate multiple compilable reference versions based on the inferred intention (as shown in \emph{Prompt 2}). 
ChatGPT then generates the reference versions (named ``\emph{Reference Version 1}'' and ``\emph{Reference Version 2}'' in  \emph{Response 2}). Note that the number of reference versions to generate is a parameter of \xxx. 
\xxx by default generates two reference versions because two is the minimal number of reference versions needed to validate the correctness of an expected output (more discussion in \S\ref{sec:dt}). Our evaluation shows that \xxx can effectively find \emph{\ft} using the default setting (\S\ref{sec:eva:rq1}).

\subsection{\ftg}\label{sec:ftg}

A \ftg is designed to find a \ftest for PUT by conducting differential testing between PUT and the reference versions generated by \pg. 
%For any test that makes PUT and its reference versions have different outputs, \xxx considers it a \ftest. 
%\emph{Component 2} in Figure~\ref{fig:workflow} shows the workflow of \ftg. Specifically, the input of \ftg is a PUT and its reference versions, the output is a \ftest candidate or a message informing \xxx's users that \xxx has not found a \ftest. 
\ftg works in three steps: \emph{generating test input for PUT} (\S\ref{sec:input}), \emph{inferring expected output using reference versions} (\S\ref{sec:dt}) and \emph{applying differential testing} (\S\ref{sec:input}).

\subsubsection{Step 1: Generating test input}\label{sec:input}

To perform differential testing, the first step is to generate test inputs that can trigger diverse behaviors (e.g., branches) of a PUT.
\ftg leverages ChatGPT to perform the generation because recent studies~\cite{deng2023large,schafer2023adaptive} reveal that LLMs have the potential to generate more diverse and valid test inputs than conventional approaches (e.g., \textsc{Pynguin}). 
Our experiment results (Figure~\ref{fig:eva:rq1buggybox}) also show that ChatGPT generates significantly more failing tests %\ft and \fc (i.e., test cases whose input can induce a failure, see \S\ref{sec:formulation} for the definitions) 
than \textsc{Pynguin}, and notably fewer illegal test inputs. Hence, \xxx prompts ChatGPT to generate diverse test inputs instead of relying on Pynguin. Note that \xxx prompts ChatGPT to ``generate diverse test input'' instead of ``generate test inputs that result in different outputs between PUT and reference versions'', because the later prompt requires ChatGPT to identify a nuance between PUT and reference versions. As shown in our illustrative example (\S\ref{sec:intro}) and our experimental result (\S\ref{sec:eva:rq3}), ChatGPT is not effective in telling the differences between two similar pieces of code.

% that \gptvalid tests generated by ChatGPT are valid, \gptvspygnuinvalid more than Pynguin~\cite{lukasczyk2022pynguin}, the state-of-the-art unit test generation tool for Python.

Step 1 works as follows. Given a PUT, \ftg requests ChatGPT to generate diverse tests for the PUT (\emph{Prompt 3} of Figure~\ref{fig:workflow}). ChatGPT then returns a set of test inputs (\emph{Response 3}). Note that the number of test inputs returned by ChatGPT may vary across conversations. In the evaluation, we repeat the experiment ten times. \ftg then chooses the test inputs in turn and performs Step 2 (\S\ref{sec:dt}) until it finds a \ftest. It discards a chosen test input if the input leads to an inconsistent output across the reference versions or cannot trigger a failure in PUT. %\scc{Also explain why not using the test inputs generated by Pynguin.}\tszon{updated in the paragraph right before this paragraph.} \scc{I don't seem to find the sentence(s) explaining why not using the test inputs from Pynguin.}\tszon{sorry that my writing may be too implicit here. In the above paragraph I pointed out that ChatGPT can generate more diverse and valid test case. Let me explicitly point out this is the reason why \xxx adopts ChatGPT to generate test input instead of Pynguin. I added ``Hence, \xxx prompts ChatGPT to generate diverse input instead of relying on Pynguin''. See if this is clearer}
%is not accepted by other steps (e.g., Step 3 only accepts test inputs that can trigger a PUT's failure, see \S\ref{sec:input}), 
%\ftg chooses the next test input and proceeds to Step 2, and so on. When all test inputs generated by ChatGPT are not accepted by other steps. \ftg requests ChatGPT to generate another set of test inputs. 

% To conduct differential testing, the first step is to generate a test input. 

% Figure~\ref{fig:workflow} shows an example of Step 1: given the PUT, \ftg requests ChatGPT to generate diverse test inputs. 
% ChatGPT then generates a set of test inputs (``$a$=12, $b$=20'',``$a$=37, $b$=60'',``$a$=11, $b$=0'' etc.). \ftg then chooses the first test input (``$a$=12, $b$=20'') and proceeds to Step 2.

% the test generation tools generate a=12 and b=20 as the test input. \todo{update the example.}

\subsubsection{Step 2: Inferring an expected output}\label{sec:dt}

In this step, \ftg infers the expected output of a PUT with respect to the test input chosen in Step 1. 
%\ftg does so by leveraging versions generated by \pg to determine the expected output. 
Since reference versions generated by ChatGPT can have bugs, they can induce incorrect failure-inducing tests. \xxx addresses this issue by using only those test inputs that lead to the same output for all reference versions. Here, the strategy assumes that the chances that all reference versions commonly suffer from the same bug are low, i.e., the reference versions are sufficiently diversified. The degree of diversity can be increased by generating more reference versions or enlarging the temperature setting of ChatGPT. Our evaluation \update{(Figure~\ref{fig:eva:rq1buggybox} and Figure~\ref{fig:eva:rq1correctbox})} by using two reference versions shows the effectiveness of this strategy: \xxx has a low probability of returning \emph{\fc} or \emph{\ff}. Essentially, \xxx returns these two types of test cases only when both reference versions commonly have the same bug. %\todo{fill in the blank}

Therefore step 2 works as follows. \ftg first passes the test input in turn to all reference versions generated by \pg. \ftg then inspects whether the output returned by each reference version is the same. If so, \ftg regards such an output as the expected output. 
%It is because intuitively, due to the stochastic nature of ChatGPT, reference versions are unlikely to suffer from the same bug. In other words, when the reference versions have the same output, such an output is likely a correct output. 
%When the reference versions have different outputs, the test input must have induced a failure from some of the reference versions. In this case, \ftg cannot infer a correct expected output. To avoid \xxx from returning an incorrect \ftest, \ftg returns to Step 1 and selects another test input. 
\ftg makes at most $k$ attempts (the default value of $k$ is 10). If \xxx cannot find a \ftest after $k$ attempts, it reports \ftests not found.

\subsubsection{Step 3: Differential testing}\label{sec:input}
% Algorithm~\ref{algo:main} shows the overall algorithm of \ftg.
In this step, \ftg inspects whether the output of PUT and the consistent output of the reference versions are the same. If not, it essentially reveals a potential failure of the PUT. Hence, when \ftg detects an output difference, \ftg constructs a \ftest with the test input found in Step 1, and the expected output inferred in Step 2, and then reports this \ftest. Otherwise, the test input is considered non-failure-inducing. %unable to induce a failure from the PUT, 
So, \ftg rolls back to Step 1 and chooses another test input (\S\ref{sec:dt}).
Before rolling back to Step 1, \ftg records the lines of code exercised by the current test input, in order to compute the branch coverage accumulated by all test inputs that have been exercised on a PUT. When the branch coverage has reached 100\% or saturated, \ftg considers a PUT has been adequately tested. Hence, \ftg stops finding a \ftest, and reports \ftests not found.

\section{Evaluation}\label{sec:evaluation}
% \subsection{Experiment setup}\label{sec:setup}
To evaluate \xxx's effectiveness and usefulness, we study four research questions. For the study of each research question, we first introduce the experiment design (e.g., subject selection, evaluation metrics, and baseline construction), followed by experimental results and discussions.

\begin{itemize} [leftmargin=*, topsep=1pt, itemsep=1pt]
  \item \textcolor{mygreen}{\faIcon{lightbulb}} \textbf{RQ1 (Finding \emph{\ft} for \quixbugs)}: \emph{Can correct \ftests for \quixbugs programs be effectively found?}
  \item \textcolor{mygreen}{\faIcon{lightbulb}} \textbf{RQ2 (Inferring program intention)}: \emph{Can program intention be effectively inferred?}
  \item \textcolor{mygreen}{\faIcon{lightbulb}} \textbf{RQ3 (Generating reference versions)}: \emph{Can reference versions be effectively generated?}
  \item \textcolor{mygreen}{\faIcon{lightbulb}} \textbf{RQ4 (Finding \emph{\ft} for \codeforces)}: \emph{Can correct \ftests for recent \codeforces programs be effectively found?}
\end{itemize}

All experiments were conducted on a Linux computer running AMD Ryzen 7 5800 8-Core Processor 3.40 GHz and 16GB RAM. All interactions with ChatGPT, such as sending requests to ChatGPT or receiving responses from ChatGPT, are performed via ChatGPT's API of version \textit{gpt-3.5-turbo-0301}. 
We conduct the experiment via ChatGPT's API instead of its web interface because ChatGPT's API allows us to explicitly specify the model version (e.g., \textit{gpt-3.5-turbo-0301}), which is useful for result reproduction. %Hence, our experiment result can be reproduced even ChatGPT is upgraded to a newer version.

% Hence, evluating \xxx on Quixbugs allows 

 % In each program, there is one bug in one line of the program's code. We chose Quixbugs for our evaluation because Quixbugs is evaluated by most LLM-based fault detection and program repair techniques \todo{citation}. 

% In addition to QuixBugs, we also apply \xxx on two popular open source project to demonstrate \xxx's usefulness on real-world projects. Specifically, we chose AAA and BBB because of some reasons \todo{fill in AAA and BBB and the reason}.

% To evaluate \xxx's effectiveness on generating failing test cases, our evaluation metrics consist of the following N \todo{N} items: Number of failing tests which trigger a true failure (NumFF), number of failing tests which trigger a coincidental failure (NumFC), number of failing tests which trigger a false failure (NumFF), number of passing tests (NumPT), number of invalid test (NumIT, which are test cases cannot be executed by a given program). 

% We implements \xxx with ChatGPT, a state-of-the-art LLM. We integrate \xxx with Evosuite for Java program, and we integrate \xxx with \textsc{Pynguin} for Python program. 

\begin{figure*}[t]
\centering
\vspace{-2mm}
\includegraphics[width=0.9\linewidth]{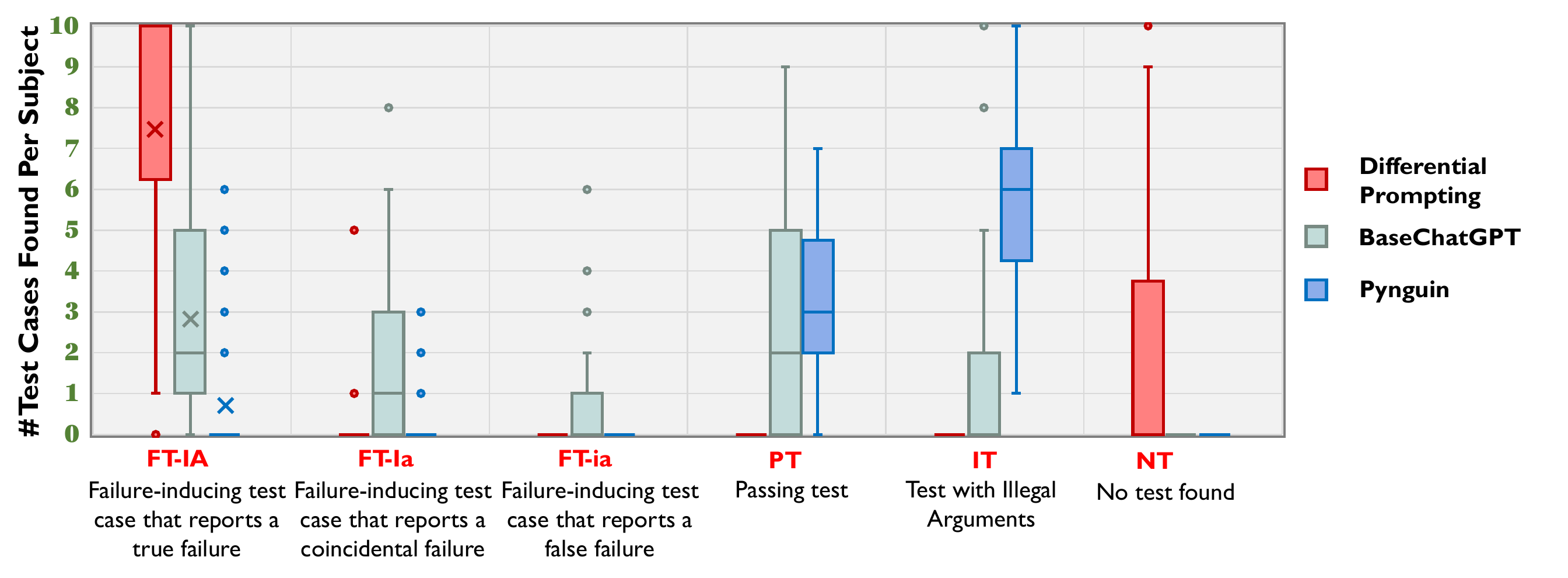}
\vspace{-1mm}
\caption{\update{Effectiveness of \xxx and the baselines in finding \ftests for buggy programs of \quixbugs. The vertical axis represents the number of test cases found by \xxx or a baseline for a program subject in ten executions. The cross marks in the \ft column indicate the average number of \ft found by the three techniques.}}
\vspace{-4mm}
%\caption{\xxx and baseline's effectiveness in generating program.}
\label{fig:eva:rq1buggybox}
\end{figure*}

\subsection{RQ1: Finding FTs for \quixbugs}\label{sec:eva:rq1}

\subsubsection{Experiment setup}\label{sec:eva:rq1:setup}
We consider two baselines to compare their effectiveness in finding \ftests:

\begin{itemize} [leftmargin=*, topsep=1pt, itemsep=1pt]
  \item \emph{\gptbaseline}: The first baseline is to prompt ChatGPT directly for \ftests. 
  This baseline prompts ChatGPT in two steps: initially, it checks with ChatGPT if a PUT contains bugs. Upon an affirmative response, it further requests ChatGPT to generate a \ftest. We refer to this baseline as \gptbaseline. Like a recent related study~\cite{sobania2023analysis}, the two-step prompting convention does not assume any knowledge if a given program is buggy, emulating a common real-life situation.

  \item \emph{\textsc{Pynguin}}: The second baseline is \textsc{Pynguin}~\cite{lukasczyk2022pynguin}, the state-of-the-art unit test generation tool for Python. 
\end{itemize}

 %\scc{After reading the writing, it seems better to call \gptbasedline as direct prompting to avoid confusion. Ok} \tszon{or BaseChatGPT? I saw Lingming's paper call the baseline this way. Ok, let's see whether it works}%, because a developer usually does not know whether a PUT contains bugs. %\scc{for the first baseline, do we also ask ChatGPT to give the expected output for each failing test inputs?}\tszon{we don't do so unless the explicit output is unknown. Usually ChatGPT will describe the expected output}
%a recently proposed prompting convention~\cite{sobania2023analysis} which queries ChatGPT in two steps (described in \S\ref{sec:intro}). 
%To the best of our knowledge, these two baselines are the state-of-the-art approaches for finding \ftests for Python programs. 

\emph{\textbf{Dataset.}} We evaluate \xxx and the baselines on \quixbugs~\cite{lin2017quixbugs}, which consists of 40 pairs of buggy and patched Python programs; each implements a commonly used algorithm, such as {\mycode breath-first-search} and {\mycode mergesort}. Each buggy program contains one bug. 
The programs have been adopted by recent works~\cite{sobania2023analysis,xia2023conversational} to study the use of LLMs for software engineering tasks. 
We select all the 80 {\mycode QuixBugs} programs as evaluation subjects because they implement algorithms that are common building blocks of real-life software~\cite{xia2023conversational,lin2017quixbugs}. %In addition, 

\update{Although \quixbugs also consists of Java programs, our evaluation focuses on Python programs for two main reasons. First, Python is one of the most popular programming languages~\cite{widyasari2020bugsinpy}. Second, recent studies~\cite{xia2023keep,xia2023conversational,liu2023your} show that ChatGPT has potential in performing various software engineering tasks for Python programs. Some of these tasks (e.g., program generation~\cite{liu2023your}) are closely related to \xxx.  Hence, the findings of these studies indicate that \xxx has potential in finding \ftests for Python programs. We leave \xxx's evaluation on Java programs to future works.}

Note that, we remove all the code comments or problem descriptions for each program documented by \quixbugs, to avoid their interference with \xxx's performance of inferring program intentions. \update{In fact, the code comments or problem descriptions of each program are considered the ground truth of the program's intention in RQ2 (\S\ref{sec:rq2}).}
%\ying{Please check the update.} \scc{Good.}

 %To evaluate \xxx's effectiveness in finding \ftests, for each buggy or correct program of {\mycode Quixbugs}, we apply \xxx to find one \ftest for the program, because in practice a developer only needs a few (e.g., one) \ftest to debug a program. 
 % This experiment design mimics a real-world scenario which a developer wants to find a failing test for a PUT. Essentially the  developer just wants one failing test for debugging (more failing test d
 % two evaluation metrics: precision and recall (discussed in \S\ref{sec:formulation}). 
 \emph{\textbf{Comparison.}} To mitigate experimentation randomness, we repeat the experiment ten times for \xxx and \gptbaseline, and record the number of \ftests in each category found. The number of test cases found each time can be either 0 or 1. %so that for each program, \xxx finds zero to ten test cases. 
For the \emph{success rate} for buggy/correct programs, we compute the ratio of number of correct \ftests found by each technique for the buggy/correct programs to the total number of times that the technique has been executed (400 = 40 programs $\times$ 10 runs). %\ying{John, please check it.}
 Unlike \xxx and \gptbaseline which return at most one test case each time, \textsc{Pynguin} returns multiple test cases. It is because \textsc{Pynguin} is a coverage-guided test generation technique designed to generate enough test cases to achieve code coverage. The current version of \textsc{Pynguin} that we can publicly access does not support parameters to limit the number of generated test cases or specify stopping conditions for test generation (e.g., coverage threshold \update{or} time budget). Hence, for a fair comparison, we repeat the experiment of applying \textsc{Pynguin} to each program ten times. In each experiment, we repeatedly execute \textsc{Pynguin} until it generates ten test cases. If \textsc{Pynguin} generates more than ten test cases, we select the top ten test cases. %As a result, \textsc{Pynguin} totally finds 100 test cases for each program.
  %Hence, for all buggy or correct programs, \textsc{Pynguin} generates four thousand test cases (forty programs times ten experiments times ten test cases). 
%We compute the number of \emph{\ft} found by \textsc{Pynguin} for each program as the number of \emph{\ft} among the one hundred test cases found by \textsc{Pynguin} divided by ten (i.e., ten experiments). In other words, 
The number of \emph{\ft} found by \textsc{Pynguin} for a program is \update{calculated as} the average number of \emph{\ft} found in the concerned ten experiments. \update{We further round off all the average numbers into integers so that the presentation of \textsc{Pynguin}'s performance is consistent with that of \xxx or \gptbaseline.}
% \update{The number has been rounded off to an integer (reason?)}. 
 %We rounded off the average number to an integer for comparison with the results of \xxx and  \gptbaseline.  
 We \update{calculate} the number of other \ftests (e.g., \emph{\fc}) found by \textsc{Pynguin} similarly. The \emph{success rate} of \textsc{Pynguin} is computed in the same way as that of \xxx and \gptbaseline. %We then adopt the same equation of computing success rate for \xxx to compute the \emph{success rate} for \textsc{Pynguin}.

\update{Apart from \emph{success rate}, we propose another evaluation metric called \emph{accuracy} to assess the three techniques' effectiveness in finding \ft. Specifically, \emph{accuracy} is calculated as the total number of \ft found by a technique for all targeted subjects (e.g., all buggy programs) divided by the total number of all test cases found by the technique for these subjects. Essentially, \emph{accuracy} is adapted from a popular evaluation metric called \emph{Precision} (i.e., a ratio calculated by diving the number of true positives with the total number of true positives and false positives). However, since \textsc{Pynguin}'s main objective is to generate test cases that achieve high code coverage instead of finding \ftests, directly using \emph{Precision} as an evaluation metric for \textsc{Pynguin} may lead to confusion. Hence, \emph{accuracy} is proposed to avoid such confusion, while sharing a similar implication with \emph{Precision}.}

 % \update{Besides \emph{success rate}, we also proposed another evaluation metric \emph{adjusted success rate}. Essentially \emph{adjusted success rate} adapts \emph{success rate} to cater for the situation where a failure-inducing input of a subject cannot be found by any of the three techniques. In other words, no FT-IA and FT-Ia test cases can be found. Adjusted success rate provides an additional comparison between the three techniques based on the programs whose failure-inducing inputs can be found in some execution.}

 % Hence, the success rate of \textsc{Pynguin} is computed as 
 
 %be branch coverage staurates 
 %we use the default setting (to the best of our knowledge, \textsc{Pynguin} 

\subsubsection{Results and findings}

\update{Figure~\ref{fig:eva:rq1buggybox}} compares the effectiveness of the three techniques in finding \ftests for buggy programs. \update{The cross marks in the \ft column indicate the average number of \ft found by the three techniques for all the forty program subjects. Particularly, the average number is calculated as the total number of \ft found for the forty subjects divided by forty.  Hence, \emph{success rate} can be calculated by dividing the average number of \ft by ten (i.e., dividing the total number of \ft by four hundred).}
% Specifically, this table shows number of FT,FF,FC,PT and IT found by \xxx and the baselines (see \S\ref{sec:formulation} for the definition of FT,FF,FC,PT and IT). 

Overall, \xxx's \emph{success rate} is \xxxft, \xxxvsgptft as \gptbaseline (\gptft) and \xxxvspynguinft as \textsc{Pynguin} (\pynguinft). \update{In terms of \emph{accuracy}, Table~\ref{tab:rq1accuracy} shows that \xxx's  accuracy for buggy code is 98.0\%, outperforms the best baseline (28.8\%) by 3.4X. This result indicates test cases returned by \xxx have a high probability to be correct \ftests (i.e., \ft).
}

%\xxx can achieve a significantly higher \emph{success rate} because \xxx has a high probability to generate correct reference versions (see \S\ref{sec:eva:rq3}), \scc{This explanation is a bit weird. On one hand, we say that reference versions do not need to be correct. On the other hand, the effectiveness of our technique is due to the correctness of reference versions. Confusing?} so that \xxx can dynamically and accurately verify a generated test's correctness (i.e., whether a generated test is \emph{\ft}). For instance, \xxx checks whether all reference versions return the same expected output with respect to the same test input (\S\ref{sec:dt}). This step allows \xxx to correctly infer the expected output of a test at a high probability. Hence, as shown in Table~\ref{tab:rq1buggy}, \xxx did not return \emph{\ff} and \emph{\iat}, and return \emph{\fc} at an extremely low probability (1.0\%). \scc{Please think about the explanation again and rewrite this paragraph. Is the high probability of correct reference versions the key that explains the effectiveness of \xxx?}

%\tszon{revised paragraph here} 
\xxx can achieve a significantly higher \emph{success rate} \update{and \emph{accuracy}} because \xxx's workflow makes use of ChatGPT's strength and bypasses its weakness. Specifically, \S\ref{sec:rq2} shows that \xxx correctly infers an intention for \correctintention buggy programs of \quixbugs, and \update{has a success rate of \xxxrv in generating reference versions that can reveal the buggy programs' failures} \S\ref{sec:eva:rq3}. 

\update{Table~\ref{tab:rq1breakdown} further delves into the three techniques' effectiveness in finding \ft}. Essentially, \xxx finds at least one \emph{\ft} in ten executions for 37 out of 40 buggy programs, and finds ten \emph{\ft} for 22 out of 40 buggy programs. In contrast, \gptbaseline (the best baseline) is restricted by ChatGPT's weakness in identifying nuance, so it finds ten \emph{\ft} only for 2 out of 40 buggy programs. \update{For \textsc{Pynguin}, it finds only few \ft. Indeed, most test cases it finds are \emph{\pt} and \emph{\iat}, and none of the \emph{\pt} consists of failure-inducing test inputs.}

\begin{table}[]
\caption{\update{Effectiveness of the three techniques in finding correct failure-inducing test cases (i.e., FT-IA) for programs of QuixBugs. Note that test cases other than FT-IA are considered incorrect failure-inducing test cases.}}
\begin{tabular}{lcc}
\hline
\rowcolor[HTML]{ECEFF4} 
\multicolumn{3}{c}{\cellcolor[HTML]{ECEFF4}Buggy programs}                                                                                                                                                                                                                                                                                                                                                                                                                                      \\ \hline
\rowcolor[HTML]{D8DBE8} 
{\color[HTML]{000000} }                                                                 & \multicolumn{1}{l}{\cellcolor[HTML]{D8DBE8}{\color[HTML]{000000} \begin{tabular}[c]{@{}l@{}}\#programs that a technique \\ finds ten FT-IA\end{tabular}}}                                     & \multicolumn{1}{l}{\cellcolor[HTML]{D8DBE8}{\color[HTML]{000000} \begin{tabular}[c]{@{}l@{}}\#programs that a technique \\ finds at least one FT-IA\end{tabular}}}                                    \\
\rowcolor[HTML]{ECEFF4} 
{\color[HTML]{000000} \begin{tabular}[c]{@{}l@{}}Differential\\ Prompting\end{tabular}} & {\color[HTML]{000000} 22}                                                                                                                                                                     & {\color[HTML]{000000} 37}                                                                                                                                                                             \\
\rowcolor[HTML]{FFFFFF} 
{\color[HTML]{000000} \begin{tabular}[c]{@{}l@{}}Base-\\ ChatGPT\end{tabular}}          & {\color[HTML]{000000} 2}                                                                                                                                                                      & {\color[HTML]{000000} 32}                                                                                                                                                                             \\
\rowcolor[HTML]{ECEFF4} 
{\color[HTML]{000000} Pynguin}                                                          & {\color[HTML]{000000} 0}                                                                                                                                                                      & {\color[HTML]{000000} 8}                                                                                                                                                                              \\
\rowcolor[HTML]{D8DBE8} 
{\color[HTML]{000000} }                                                                 & \multicolumn{1}{l}{\cellcolor[HTML]{D8DBE8}{\color[HTML]{000000} \begin{tabular}[c]{@{}l@{}}\#programs that a technique \\ finds ten incorrect failure-\\ inducing test cases\end{tabular}}}  & \multicolumn{1}{l}{\cellcolor[HTML]{D8DBE8}{\color[HTML]{000000} \begin{tabular}[c]{@{}l@{}}\#programs that a technique \\ finds at least one incorrect \\ failure-inducing test cases\end{tabular}}} \\
\rowcolor[HTML]{ECEFF4} 
{\color[HTML]{000000} \begin{tabular}[c]{@{}l@{}}Differential\\ Prompting\end{tabular}} & {\color[HTML]{000000} 0}                                                                                                                                                                      & {\color[HTML]{000000} 2}                                                                                                                                                                              \\
\rowcolor[HTML]{FFFFFF} 
{\color[HTML]{000000} \begin{tabular}[c]{@{}l@{}}Base-\\ ChatGPT\end{tabular}}          & {\color[HTML]{000000} 8}                                                                                                                                                                      & {\color[HTML]{000000} 38}                                                                                                                                                                             \\
\rowcolor[HTML]{ECEFF4} 
{\color[HTML]{000000} Pynguin}                                                          & {\color[HTML]{000000} 32}                                                                                                                                                                     & {\color[HTML]{000000} 40}                                                                                                                                                                             \\ \hline
\rowcolor[HTML]{ECEFF4} 
\multicolumn{3}{c}{\cellcolor[HTML]{ECEFF4}{\color[HTML]{000000} Correct programs}}                                                                                                                                                                                                                                                                                                                                                                                                             \\ \hline
\rowcolor[HTML]{D8DBE8} 
{\color[HTML]{000000} }                                                                 & \multicolumn{1}{l}{\cellcolor[HTML]{D8DBE8}{\color[HTML]{000000} \begin{tabular}[c]{@{}l@{}}\#programs that a technique \\ finds ten incorrect failure- \\ inducing test cases\end{tabular}}} & \multicolumn{1}{l}{\cellcolor[HTML]{D8DBE8}{\color[HTML]{000000} \begin{tabular}[c]{@{}l@{}}\#programs that a technique \\ finds at least one incorrect \\ failure-inducing test cases\end{tabular}}} \\
\rowcolor[HTML]{ECEFF4} 
{\color[HTML]{000000} \begin{tabular}[c]{@{}l@{}}Differential\\ Prompting\end{tabular}} & {\color[HTML]{000000} 0}                                                                                                                                                                      & {\color[HTML]{000000} 2}                                                                                                                                                                              \\
\rowcolor[HTML]{FFFFFF} 
{\color[HTML]{000000} \begin{tabular}[c]{@{}l@{}}Base-\\ ChatGPT\end{tabular}}          & {\color[HTML]{000000} 36}                                                                                                                                                                     & {\color[HTML]{000000} 40}                                                                                                                                                                             \\
\rowcolor[HTML]{ECEFF4} 
{\color[HTML]{000000} Pynguin}                                                          & {\color[HTML]{000000} 40}                                                                                                                                                                     & {\color[HTML]{000000} 40}                                                                                                                                                                             \\ \hline
\end{tabular}
\label{tab:rq1breakdown}
\end{table}

\begin{table}[]
\centering
\caption{\update{Accuracy of the three techniques on \quixbugs programs and \codeforces programs. Accuracy is calculated as the number of \ft divided by the number of all test cases (\S\ref{sec:eva:rq1:setup}). It has a similar implication as \emph{Precision}. }}
\begin{tabular}{lcc}
\hline
\rowcolor[HTML]{ECEFF4} 
                                                                                        & \multicolumn{1}{l}{\cellcolor[HTML]{ECEFF4}\begin{tabular}[c]{@{}l@{}}Accuracy (buggy \\ programs only)\end{tabular}} & \multicolumn{1}{l}{\cellcolor[HTML]{ECEFF4}\begin{tabular}[c]{@{}l@{}}Accuracy (buggy and \\ correct programs)\end{tabular}} \\ \hline
\rowcolor[HTML]{D8DBE8} 
\multicolumn{3}{c}{\cellcolor[HTML]{D8DBE8}QuixBugs programs}                                                                                                                                                                                                                                                                                  \\
\rowcolor[HTML]{ECEFF4} 
{\color[HTML]{000000} \begin{tabular}[c]{@{}l@{}}Differential\\ Prompting\end{tabular}} & {\color[HTML]{000000} 98.0\%}                                                                                         & {\color[HTML]{000000} 94.6\%}                                                                                                \\
\rowcolor[HTML]{FFFFFF} 
BaseChatGPT                                                                             & 28.8\%                                                                                                                & 14.3\%                                                                                                                       \\
\rowcolor[HTML]{ECEFF4} 
Pynguin                                                                                 & 7.5\%                                                                                                                 & 3.8\%                                                                                                                        \\
\rowcolor[HTML]{D8DBE8} 
\multicolumn{3}{c}{\cellcolor[HTML]{D8DBE8}Codeforces programs}                                                                                                                                                                                                                                                                                \\
\rowcolor[HTML]{ECEFF4} 
\begin{tabular}[c]{@{}l@{}}Differential\\ Prompting\end{tabular}                        & 87.9\%                                                                                                                & 80.1\%                                                                                                                       \\
\rowcolor[HTML]{FFFFFF} 
{\color[HTML]{333333} BaseChatGPT}                                                      & {\color[HTML]{333333} 7.1\%}                                                                                          & {\color[HTML]{333333} 3.6\%}                                                                                                 \\
\rowcolor[HTML]{ECEFF4} 
Pynguin                                                                                 & 0.0\%                                                                                                                 & 0.0\%                                                                                                                       
\end{tabular}
\label{tab:rq1accuracy}
\vspace{-7mm}
\end{table}

We analyze the few cases where \xxx fails to find \ftests successfully and observe two situations.
The \emph{first situation} occurs when \xxx cannot find a failure-inducing test input. For instance, \xxx finds \emph{\ft} in only three executions for {\mycode depth-first-search}. In the remaining seven executions, \xxx cannot generate a test input that reaches the buggy branch using ChatGPT. Specifically, to reach the buggy branch, a test input needs to contain a graph of at least one cycle. Without guidance, \xxx (and also \gptbaseline) does not always generate such a test input. This situation also occurs in the other three buggy programs {\mycode detect-cycle}, {\mycode quick-sort}, and {\mycode topological-ordering}, where \xxx accomplishes a relatively low \emph{success rate}. 
%,  %{\mycode subsequences} and  {\mycode topological-ordering}. 
%because \xxx cannot find a test input that covers the PUTs' buggy branch. 
%\scc{Add a couple of sentences if \textsc{Pynguin} can generate test inputs reaching the buggy branches of these subjects? And if possible, a sentence explaining why \textsc{Pynguin} that uses a coverage-guided technique still cannot generate the inputs to reach the buggy branches.}\tszon{working on it} A possible solution is to guide ChatGPT to generate more diverse test inputs, see \S\ref{sec:discussion} for discussion.

The  \emph{second situation} occurs when the two reference versions generated by ChatGPT suffer from the same bug. As a result, \xxx can return incorrect \ftests (e.g., {\mycode lcs-length} and {\mycode wrap} return two and one \emph{\fc} respectively). \S\ref{sec:eva:rq3} further delves into this reason and methods for mitigation.

\begin{tcolorbox}[boxrule=1pt,boxsep=1pt,left=2pt,right=2pt,top=2pt,bottom=2pt]
% \small
 \textcolor{mygreen}{\faIcon{key}} \noindent\emph{\textbf{Finding 1:}} \emph{For buggy programs of \quixbugs, \xxx's success rate is \xxxft, \xxxvsgptft as \gptbaseline (\gptft) and \xxxvspynguinft as Pynguins (\pynguinft).
 }
	
 \textcolor{red}{\faIcon{user-edit}} \noindent\emph{\textbf{Implication:}} \emph{\xxx is effective in finding \ftests for buggy programs.
 }
\end{tcolorbox} 

% Specifically, we found that \xxx becomes less accurate in generating reference versions when a PUT has self-defined object (e.g., a tree object). 
% For instance, the original code of {\mycode topological\_ordering} constructs a tree object of four nodes, yet the reference versions generated by \xxx constructs a tree object of only three nodes. {\mycode shortest\_path\_length} also has a similar problem. We further elaborate on this reason in RQ3 (\S\ref{sec:eva:rq3}). Fortunately, even reference versions generated by \xxx are incorrect, \xxx's \ftg prevents these incorrect reference versions from causing \xxx to return incorrect tests.

% A final reason is that \xxx inferred an intention incorrectly, which makes \xxx generate an incorrect reference versions. For instance, for the subject {\mycode Subsequences}, the actual intention should be \textit{Generate all possible sequences of integers from a to b (inclusive)}, yet the intention inferred by \xxx is \textit{Generall all possible sequences of integers from a to b (exclusive)}. We further elaborate on this reason in RQ3 (\S\ref{sec:rq3:finding}).

% \input{table/rq1correct}

\begin{figure}[t]
\centering
\vspace{-2mm}
\includegraphics[width=0.9\linewidth]{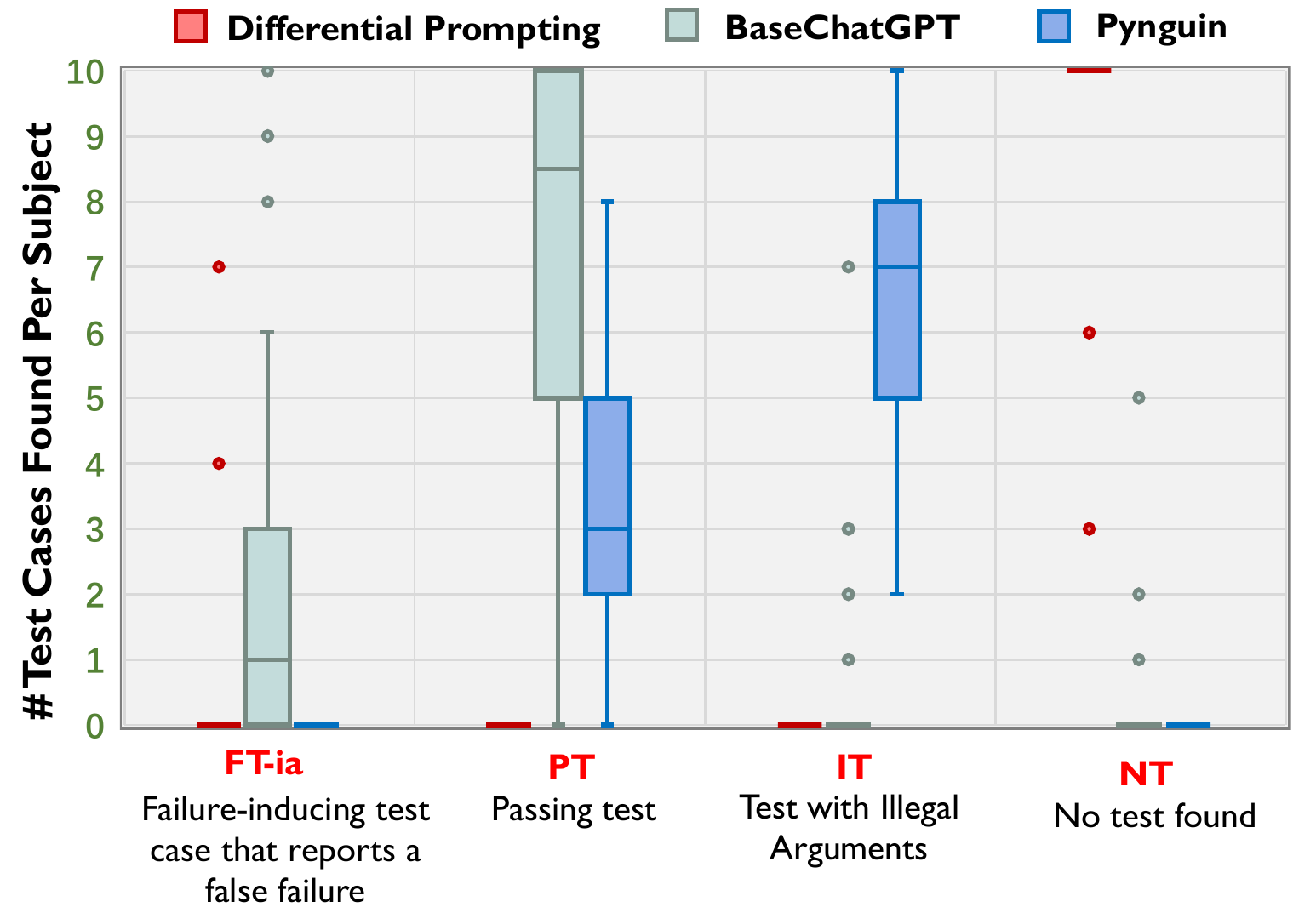}
\vspace{-1mm}
\caption{\update{Effectiveness of \xxx and the baselines in finding \ftests for correct programs of \quixbugs. The vertical axis represents the number of test cases found by \xxx or a baseline for a program subject in ten executions.}} %\scc{Try to compress it to single column.}} 
\vspace{-4mm}
%\caption{\xxx and baseline's effectiveness in generating program.}
\label{fig:eva:rq1correctbox}
\end{figure}

\update{Figure~\ref{fig:eva:rq1correctbox} compares the effectiveness of \xxx and the baselines on correct programs of \quixbugs. \update{The figure shows that \xxx has notably lower frequencies in finding incorrect \ftests (i.e., \emph{\ff}, \emph{\pt}, and \emph{\iat}).} Table~\ref{tab:rq1breakdown} further looks into the three techniques' performance in avoiding incorrect \ftests for correct programs. Essentially, \xxx returns incorrect \ftests for only two programs. These two programs are {\myfont lcs-length} and {\myfont wrap}, which \xxx also returns incorrect \ftests for their buggy version (\S\ref{sec:rq2:finding} discusses these two cases). Meanwhile, the baselines return incorrect \ftests for all the forty programs. }

\update{Overall, Table~\ref{tab:rq1accuracy} shows that \xxx's accuracy on both buggy and correct programs (94.6\%) is comparable with that on buggy programs only (98.0\%). Moreover, \xxx's accuracy on buggy and correct programs outperforms the best baseline (14.3\%) by 6.6X.}

% \update{In terms of accuracy}

%Besides, the probability of \xxx returns an incorrect \ftest (i.e., for the subjects {\mycode lcs-length} and {\mycode wrap}) is  low (0\%-\xxxffcorrect). \S\ref{sec:eva:rq3} delves into the reason and methods for mitigation.

% The reason that \xxx return an incorrect \ftest for correct programs is similar as the reason that \xxx returns incorrect \ftest for buggy programs
% The reason is similar 

\begin{tcolorbox}[boxrule=1pt,boxsep=1pt,left=2pt,right=2pt,top=2pt,bottom=2pt]
% \small
 \textcolor{mygreen}{\faIcon{key}} \noindent\emph{\textbf{Finding 2:}} \update{\emph{
 % For correct programs, \xxx returns incorrect \ftests for only two programs. In contrast, the baselines return incorrect \ftests for all the forty programs. Overall, 
 \xxx's accuracy on buggy and correct programs (94.6\%) outperforms the best baseline (14.3\%) by 6.6X.}}\\
 \textcolor{red}{\faIcon{user-edit}} \noindent\emph{\textbf{Implication:}} \update{\emph{\xxx has a high probability of returning correct \ftests (i.e., returning \ft for buggy programs, and returning no \ftest for correct programs).}
 }
\end{tcolorbox}

\subsection{RQ2: Inferring Program Intention}\label{sec:rq2}

\subsubsection{Experiment setup}

To evaluate the likelihood that ChatGPT is able to infer a program's intention correctly even if the program is buggy, we conduct a manual analysis. %s effectiveness in inferring program intention, we study the number of intentions correctly inferred by \xxx. \scc{Should this RQ evaluates the validity of the observation that ChatGPT is able to infer the intention of a given program? We are not evaluating \xxx? This is why baselines are not needed?}\tszon{Yes, this RQ is an ablation study, so we evaluate ChatGPT instead of \xxx, and do not construct a baseline} 
We treat the problem description documented in the comments of each program in \quixbugs as the ground truth to assess the correctness of the 400 intentions inferred by ChatGPT in RQ1. 
Following an open coding procedure~\cite{open-coding}, 
%We analyze the 400 intentions inferred by ChatGPT in RQ1. %\xxx has been applied to each buggy program ten times, hence \xxx has inferred ten intentions for each program. We collect these intentions and manually investigate whether each intention is correct. Specifically, 
two authors who have substantial software
development experience separately examine whether each inferred intention is consistent with the documented problem description. Discrepancies between the two authors are discussed until a consensus is reached. \update{Upon completion of the manual inspection, we observed Cohen's Kappa to be 82.3\%, which is considered an ``almost perfect agreement'' by Landis et al.~\cite{landis1977measurement}. We then calculate ChatGPT's success rate in inferring intentions as the number of correct intentions divided by 400 (i.e., total number of intention returned by ChatGPT in RQ1).}

%\scc{Please check if my writing is correct.}
%\ying{Please check the update.} \scc{Good.} %\scc{We should follow the open coding methodology.} %Then, these two students reach a consensus under the moderation o©f the students' supervisor. 
% \xxx's effectiveness in inferring code intention is built upon ChatGPT's effectiveness in inferring code intention (\S\ref{sec:pg}). Hence, to answer this RQ, we request ChatGPT to infer a program's intention 
% To analyze the reason behind \xxx's effectiveness, we evaluate the effectiveness \xxx's three core components: \encoder, \decoder and \generator. 

% To evaluate the effectiveness of \encoder, we investigate \encoder's accuracy in summarizing a program's intention. To do so, we conduct a manual check with two people. 

% To evaluate the effectiveness of \generator, we investigate how many steps are required to

\subsubsection{Results and findings}\label{sec:rq2:finding}

\begin{figure}[t]
\centering
\includegraphics[width=0.55\linewidth]{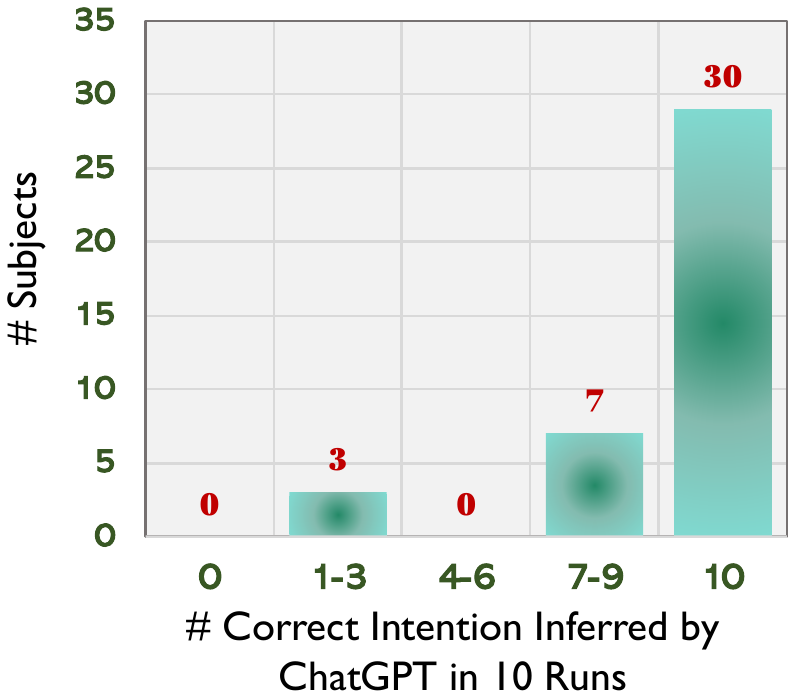}
\vspace{-1mm}
\caption{ChatGPT's effectiveness in inferring program intention.}
\vspace{-4mm}
\label{fig:eva:intention}
\end{figure}

\update{Figure~\ref{fig:eva:intention} shows that ChatGPT infers a correct intention for a majority of buggy programs. Specifically, \xxx correctly infers ten intentions for 30 out of 40 subjects. Overall, \xxx's success rate in inferring intention is \correctintention (\numcorrectintention out of 400).
This result justifies our insight that ChatGPT is good at inferring program intention despite the presence of program bugs. }

% Figure~\ref{fig:eva:correctintention} shows an example of a correct intention inferred by \xxx for a buggy {\mycode gcd} program. Essentially, the {\mycode gcd} program mistakenly  
% \textit{divides a by b until a becomes 0}, while the intention inferred by ChatGPT(Figure~\ref{fig:eva:correctintention}) correctly points out that \textit {the function recursively divides a by b until b becomes 0}. This example shows that \xxx is insensitive to the presence of nuance when inferring the {\mycode gcd}'s intention.

%\tszon{update} 
We further look into the three subjects ({\mycode lcs-length}, {\mycode wrap} and {\mycode next\_palindrome}) where \xxx infers only few (1-3) correct intentions. Specifically, the intentions inferred by \xxx for these subjects often describe the actual program semantics of the incorrect implementation. However, these intentions are incorrect, missing important details about the program's intended functionality. 

%\tszon{update} 
\update{For instance, {\mycode lcs-length} aims to solve the problem of finding the longest common substring of two input strings (``common substrings'' are defined as consecutive characters that exist in both input strings). The buggy PUT intends to address this problem using dynamic programming. However, the PUT implements an incorrect dynamic programming algorithm: the implemented statement is \texttt{dp[i,j]=dp[i-1,j]+1}, while the correct statement should be \texttt{dp[i,j]=dp[i-1,j-1]+1}.
Hence, the PUT often finds common characters that are not the longest, if not the shortest.}

\update{The intention inferred by ChatGPT correctly points out that the PUT implements a dynamic programming algorithm for finding the longest common characters. %\update{and correctly points out the dynamic programming is implemented with \texttt{dp[i,j] = dp[i-1,j-1]+1}}. 
However, the intention does not mention whether the characters should be consecutive or not. Hence, the reference versions generated based on this intention often output the longest common \update{characters} that are not consecutive (i.e., these reference versions are bad). A similar situation occurs for {\mycode wrap} and {\mycode next\_palindrome}. \S\ref{sec:eva:rq3} provides further discussion about this case and a possible mitigation strategy.}

\begin{tcolorbox}[boxrule=1pt,boxsep=1pt,left=2pt,right=2pt,top=2pt,bottom=2pt]
% \small
 \textcolor{mygreen}{\faIcon{key}} \noindent\emph{\textbf{Finding 3:}} \emph{\update{\xxx's success rate in inferring intention is \correctintention (\numcorrectintention out of 400).}
 }
	
 \textcolor{red}{\faIcon{user-edit}} \noindent\emph{\textbf{Implication:}} \emph{ChatGPT can often infer the intention of a program despite the presence of bugs. % ignores the presence of a bug (i.e., nuances) when inferring a buggy program's intention.
 }
\end{tcolorbox}

\subsection{RQ3: Generating Reference Versions}\label{sec:eva:rq3}

\subsubsection{Experiment setup}

To evaluate \xxx's effectiveness in generating good reference versions, \update{we study the number of good reference versions generated by \xxx: a reference version is considered \emph{good} if the reference version does not suffer from the same bug(s) as the PUT (discussed in \S\ref{sec:pg:overview}).}

%when the reference version passes all test cases provided by {\mycode Quixbugs} for testing a PUT.

Specifically, recalled that in RQ1, \xxx has been applied to each buggy program 10 times. Each time, \xxx generates two reference versions (\S\ref{sec:pg:worflow}). Hence, for each buggy program, \xxx has generated 20 reference versions in total. We consider a reference version of a buggy program to be good if it passes the failure-inducing test cases provided by {\mycode QuixBugs} for the program. %These test cases basically include failure-inducing test cases of a PUT.
%If so, a reference version is considered correct, otherwise a reference version is considered incorrect. 

In this RQ, \xxx's \textbf{\emph{baseline}} is a strawman approach that directly \update{prompts} ChatGPT to generate reference versions. 
The strawman approach prompts ChatGPT in two steps. 
First, it asks ChatGPT whether a PUT has bugs. Upon affirmative response, it further asks ChatGPT to generate two bug-fixed implementations of the PUT (\update{same as} \xxx that in each execution, \xxx also generates two reference versions). This two-step prompting convention emulates a common real-life scenario (similar to \gptbaseline introduced in RQ1). 
Following our experiment setup in RQ1, we apply the strawman approach to each program 10 times. 
Our \textbf{\emph{evaluation metric}} for this RQ is \update{\emph{success rate} in generating good reference versions: the \emph{success rate} is calculated as the number of good reference versions generated by a technique for the forty buggy programs divided by eight hundred (which is the total number of reference versions generated by the technique for the forty buggy programs).}

% \emph{the number of good reference versions found by \xxx or the baseline for each subject.} 
%Hence, for each buggy

% Since \xxx generates two reference versions for each subject, while the baseline generates one reference version for each subject, we divide the number of correct reference versions generated by \xxx by two. By doing so, the result of \xxx and the baseline are comparable.

\subsubsection{Results and findings}\label{sec:rq3:finding}

\begin{figure}[t]
\centering
\vspace{-2mm}
\includegraphics[width=0.8\linewidth]{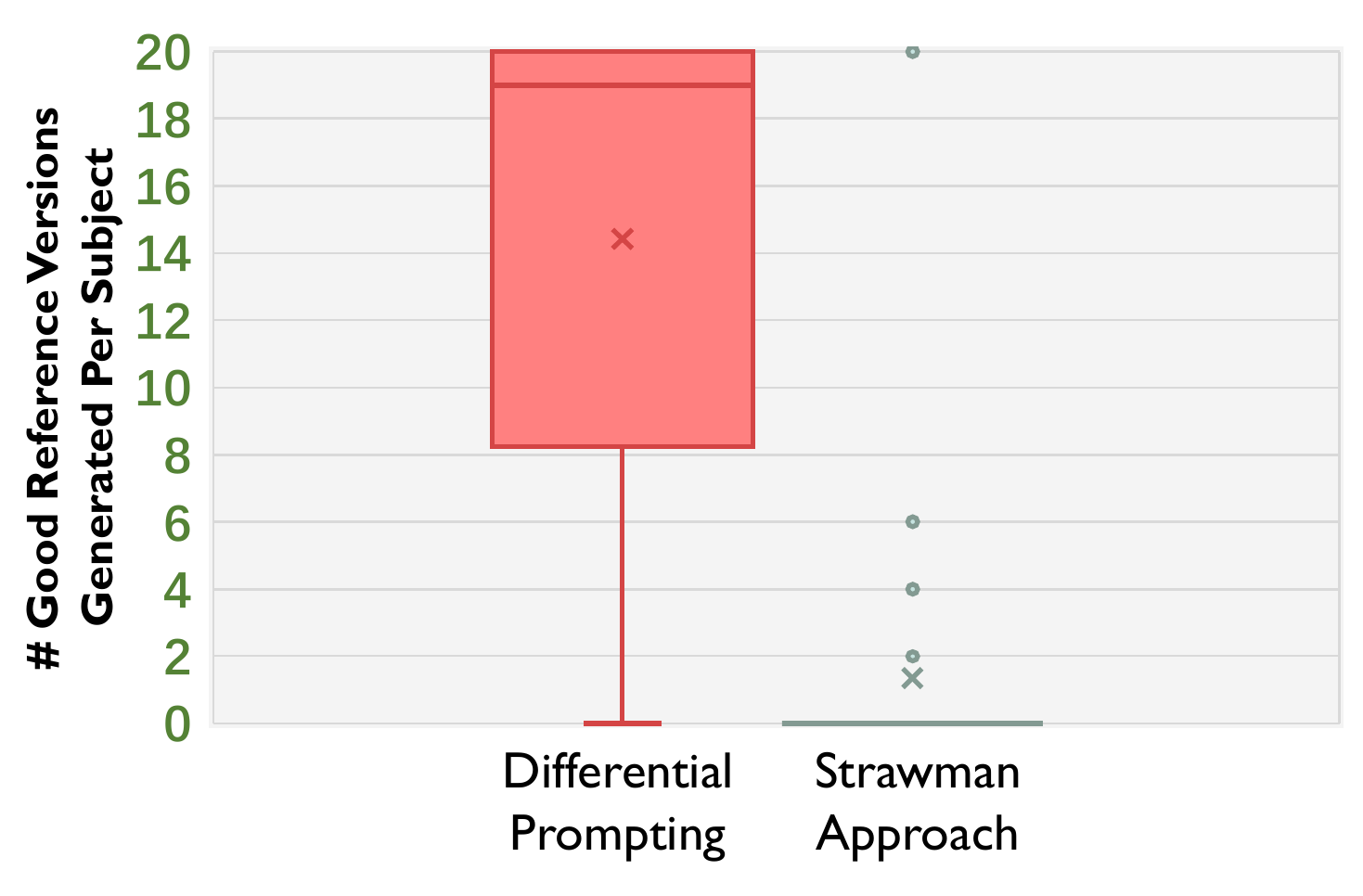}
\vspace{-3mm}
\caption{\update{Effectiveness of \xxx and the baseline in generating good reference versions. The vertical axis represents the number of good reference versions generated by \xxx or the baseline for a subject in ten executions. Note that \xxx or the baseline generates two reference versions in each execution. The cross marks in the figure indicate the average number of good reference versions generated by the two techniques.} %\scc{Update the label of horizontal axis from Effective to Good. Move the Differential Prompting and ChatGPT box to the right of the bar chart or top right hand corner inside the bar chart. It does not need to be boxed. Change ChatGPT to Strawman Approach.}
%\scc{This diagram is difficult to follow. Please explain its meaning in the caption. Better find an alternative means to present it.}\tszon{Well noted. Will find an alternative way to plot it.} 
}
\vspace{-4mm}
%\caption{\xxx and baseline's effectiveness in generating program.}
\label{fig:eva:generate}
\end{figure}

\update{Figure~\ref{fig:eva:generate} compares the effectiveness of \xxx and the baseline in generating reference versions. The cross marks in the figure indicate the average number of good reference versions generated by the two techniques for all the forty program subjects. Specifically, the average number is calculated by dividing the total number of good reference versions by forty. Hence, the \emph{success rate} can be calculated by dividing the average number by twenty  (i.e., dividing the total number of good reference versions by eight hundred).
}

Overall, \xxx's success rate is \xxxrv, outperforms the baseline (\gptrv) by \xxxrvgptrv. Besides, \xxx generates 20 good reference versions for 20 subjects, while the baseline generates 20 good reference versions for only two subjects. It is because \xxx generates reference versions through a PUT's intention (instead of its implementation), and hence bypasses ChatGPT's weakness in identifying nuances(\S\ref{sec:rq2:finding}). %Since the intention does not consist of a PUT's bug (\S\ref{sec:rq2:finding}), \xxx has a high probability to generate a correct reference version.
In contrast, the strawman approach generates reference versions from a PUT's implementation, hence it confronts ChatGPT's limitation in identifying nuances. \update{Figure~\ref{fig:eva:generate} shows that the strawman approach cannot generate any good reference version for a majority of (34 out of 40) subjects.} Specifically, in \chatgptnobugfound of the executions using the strawman approach, ChatGPT responds with ``\emph{no bug is found}''. In \chatgptmoreinformationrequired of the executions, ChatGPT cannot confirm whether a bug exists in a PUT and responds with ``\emph{More information is required}''. The result is in line with our conjecture that ChatGPT is insensitive to nuances (e.g., bugs) in input tokens (\S\ref{sec:intro}).

For the subjects that \xxx only generates a low number (e.g., nearly zero) of good reference versions (e.g.,  {\mycode lcs\_length}, {\mycode next\_palindrome}, {\mycode wrap}), we found the main reason is that \xxx has inferred an incorrect intention (\S\ref{sec:rq2}). For instance, 
the intention inferred by \xxx for {\mycode lcs\_length} often misses important details, causing \xxx to generate reference versions that always output an incorrect value. Hence, for {\mycode lcs\_length}, \xxx generates many (18 out of 20) bad reference versions.

However, we found that by providing \xxx with the correct intention (i.e., the documented description). \xxx has a much higher probability of generating a good reference version. For instance, with a correct intention, the number of good reference versions for {\mycode lcs\_length} generated by \xxx increases from 2 to 20 out of 20. We also observed a similar increase for {\mycode next\_palindrome} and {\mycode wrap}. This result essentially implies inferring a correct intention is crucial. \update{One possible solution is adopting in-context learning (e.g., few-shot prompting), which has shown great potential in improving ChatGPT's reasoning capability.~\cite{feng2023prompting}}

\begin{tcolorbox}[boxrule=1pt,boxsep=1pt,left=2pt,right=2pt,top=2pt,bottom=2pt]
% \small
 \textcolor{mygreen}{\faIcon{key}} \noindent\emph{\textbf{Finding 4:}} \update{\emph{ \xxx's success rate in generating good reference versions is \xxxrv, outperforms the baseline (\gptrv) by \xxxrvgptrv. 
 % Besides, \xxx generates 20 good reference versions for 20 subjects, while the baseline generates 20 good reference versions for only 2 subjects.
 }}
	
 \textcolor{red}{\faIcon{user-edit}} \noindent\emph{\textbf{Implication:}} \emph{\update{\xxx is effective in generating good reference versions.}
 % XXX
 }
\end{tcolorbox}

\begin{table*}[]
  \caption{The effectiveness of Differential Prompting and the baselines in finding failure-inducing test cases for \codeforces.
  %\scc{How may readers interpret the colors in the table? Looks confusing. I cannot see the color differences among 3, 4, 2, and 5, 6, 7}
  }
 %\renewcommand{\arraystretch}{1.2}
  %   \resizebox{1.0\linewidth}{!}{
    \scriptsize
    \centering
     \setlength\tabcolsep{0.1pt}     % horizental space
     \def\arraystretch{1}
     \vspace{-2mm}
     \bgroup
  
 \begin{tabular}{|lccccccccccccccc|}
 \toprule
 \multicolumn{ 1}{|l|}{}                             & \multicolumn{ 3}{c|}{\begin{tabular}[c]{@{}c@{}}\textbf{\emph{Failing  test which}}\\       \textbf{\emph{triggers a true failure}} (\emph{\ft})\end{tabular}}                                                                                                                               & \multicolumn{ 3}{c|}{\begin{tabular}[c]{@{}c@{}}\textbf{\emph{Failing test which triggers}} \\      \textbf{\emph{a coincidental failure}} (\emph{\fc})\end{tabular}}                                                                                                                     & \multicolumn{ 3}{c|}{\begin{tabular}[c]{@{}c@{}}\textbf{\emph{Failing test which triggers}} \\ \textbf{\emph{a false failure}} (\emph{\ff})\end{tabular}}                                                                                                     & \multicolumn{ 3}{c|}{\textbf{\emph{Passing test}} (\emph{\pt})}                                                                                                                                                                                    & \multicolumn{ 3}{c|}{\textbf{\emph{Illegal test input}} (\emph{\iat})}                                                                                                                                                          \\ \hline
 \multicolumn{ 1}{|c|}{\textbf{\emph{Buggy programs}}}                             & \multicolumn{ 1}{c|}{\cellcolor[HTML]{F19999}{\color[HTML]{000000} \begin{tabular}[c]{@{}c@{}}\emph{Differential} \\ \emph{Prompting}\end{tabular}}} & \multicolumn{ 1}{c|}{\cellcolor[HTML]{83CCBC}\begin{tabular}[c]{@{}c@{}}\emph{Base-} \\ \emph{ChatGPT}\end{tabular}} & \multicolumn{ 1}{c|}{\cellcolor[HTML]{6cc6f8}\emph{\textsc{Pynguin}}} & \multicolumn{ 1}{c|}{\cellcolor[HTML]{F19999}{\color[HTML]{000000} \begin{tabular}[c]{@{}c@{}}\emph{Differential}\\ \emph{Prompting}\end{tabular}}} & \multicolumn{ 1}{c|}{\cellcolor[HTML]{83CCBC}\begin{tabular}[c]{@{}c@{}}\emph{Base-} \\ \emph{ChatGPT}\end{tabular}} & \multicolumn{ 1}{c|}{\cellcolor[HTML]{6cc6f8}\emph{\textsc{Pynguin}}} & \multicolumn{ 1}{c|}{\cellcolor[HTML]{F19999}\begin{tabular}[c]{@{}c@{}}\emph{Differential}\\ \emph{Prompting}\end{tabular}} & \multicolumn{ 1}{c|}{\cellcolor[HTML]{83CCBC}\begin{tabular}[c]{@{}c@{}}\emph{Base-} \\ \emph{ChatGPT}\end{tabular}} & \multicolumn{ 1}{c|}{\cellcolor[HTML]{6cc6f8}\emph{\textsc{Pynguin}}} & \multicolumn{ 1}{c|}{\cellcolor[HTML]{F19999}\begin{tabular}[c]{@{}c@{}}\emph{Differential}\\ \emph{Prompting}\end{tabular}} & \multicolumn{ 1}{c|}{\cellcolor[HTML]{83CCBC}\begin{tabular}[c]{@{}c@{}}\emph{Base-} \\ \emph{ChatGPT}\end{tabular}} & \multicolumn{ 1}{c|}{\cellcolor[HTML]{6cc6f8}\emph{\textsc{Pynguin}}} & \multicolumn{ 1}{c|}{\cellcolor[HTML]{F19999}\begin{tabular}[c]{@{}c@{}}\emph{Differential}\\ \emph{Prompting}\end{tabular}} & \multicolumn{ 1}{c|}{\cellcolor[HTML]{83CCBC}\begin{tabular}[c]{@{}c@{}}\emph{Base-} \\ \emph{ChatGPT}\end{tabular}} & \cellcolor[HTML]{6cc6f8}\emph{\textsc{Pynguin}} \\ \hline
 \multicolumn{16}{|c|}{\textbf{\emph{Buggy program}}}                                                                                                                                                                                                                                                                                                                                                                                                                                                                                                                                                                                                                                                                                                                                                                                                                                                                                                                                                                                                                                                                                                                                                                                               \\ \hline
 \multicolumn{ 1}{|l|}{\emph{\textbf{A:}} Medium Number}                & \multicolumn{ 1}{c|}{\cellcolor[HTML]{FFFFE0}  10}                                                                                                               & \multicolumn{ 1}{c|}{\cellcolor[HTML]{FFFFE0} 4}                               & \multicolumn{ 1}{c|}{\cellcolor[HTML]{FFFFE0} 0}                               & \multicolumn{ 1}{c|}{\cellcolor[HTML]{FFFFE0}  0}                                                                                                               & \multicolumn{ 1}{c|}{\cellcolor[HTML]{FFFFE0} 2}                               & \multicolumn{ 1}{c|}{\cellcolor[HTML]{FFFFE0}  0}                               & \multicolumn{ 1}{c|}{\cellcolor[HTML]{FFFFE0}  0}                                                                                        & \multicolumn{ 1}{c|}{\cellcolor[HTML]{FFFFE0}  1}                               & \multicolumn{ 1}{c|}{\cellcolor[HTML]{FFFFE0}  0}                               & \multicolumn{ 1}{c|}{\cellcolor[HTML]{FFFFE0}  0}                                                                                        & \multicolumn{ 1}{c|}{\cellcolor[HTML]{FFFFE0} 2}                               & \multicolumn{ 1}{c|}{\cellcolor[HTML]{FFFFE0} 2}                               & \multicolumn{ 1}{c|}{\cellcolor[HTML]{FFFFE0}  0}                                                                                        & \multicolumn{ 1}{c|}{\cellcolor[HTML]{FFFFE0}  1}                               & {\cellcolor[HTML]{FFFFE0} 8}                               \\ \hline
 \multicolumn{ 1}{|l|}{\emph{\textbf{B:}} Atilla's   Favorite}  & \multicolumn{ 1}{c|}{\cellcolor[HTML]{FFFFE0} 6}                                                                                                                & \multicolumn{ 1}{c|}{\cellcolor[HTML]{FFFFE0} 1}                               & \multicolumn{ 1}{c|}{\cellcolor[HTML]{FFFFE0} 0}                                                            & \multicolumn{ 1}{c|}{\cellcolor[HTML]{FFFFE0}  0}                                                                                                               & \multicolumn{ 1}{c|}{\cellcolor[HTML]{FFFFE0} 5}                               & \multicolumn{ 1}{c|}{\cellcolor[HTML]{FFFFE0}  0}                               & \multicolumn{ 1}{c|}{\cellcolor[HTML]{FFFFE0}  0}                                                                                        & \multicolumn{ 1}{c|}{\cellcolor[HTML]{FFFFE0} 3}                               & \multicolumn{ 1}{c|}{\cellcolor[HTML]{FFFFE0}  0}                               & \multicolumn{ 1}{c|}{\cellcolor[HTML]{FFFFE0}  0}                                                                                        & \multicolumn{ 1}{c|}{\cellcolor[HTML]{FFFFE0}  0}                               & \multicolumn{ 1}{c|}{\cellcolor[HTML]{FFFFE0} 2}                               & \multicolumn{ 1}{c|}{\cellcolor[HTML]{FFFFE0}  0}                                                                                        & \multicolumn{ 1}{c|}{\cellcolor[HTML]{FFFFE0}  1}                               & {\cellcolor[HTML]{FFFFE0} 8}                               \\ \hline
 \multicolumn{ 1}{|l|}{\emph{\textbf{C:}} Advantage}                    & \multicolumn{ 1}{c|}{\cellcolor[HTML]{FFFFE0} 4}                                                                                                                & \multicolumn{ 1}{c|}{\cellcolor[HTML]{FFFFE0} 0}                               & \multicolumn{ 1}{c|}{\cellcolor[HTML]{FFFFE0} 0}                                                    & \multicolumn{ 1}{c|}{\cellcolor[HTML]{FFFFE0}  0}                                                                                                               & \multicolumn{ 1}{c|}{\cellcolor[HTML]{FFFFE0} 5}                               & \multicolumn{ 1}{c|}{\cellcolor[HTML]{FFFFE0}  0}                               & \multicolumn{ 1}{c|}{\cellcolor[HTML]{FFFFE0}  0}                                                                                        & \multicolumn{ 1}{c|}{\cellcolor[HTML]{FFFFE0} 4}                               & \multicolumn{ 1}{c|}{\cellcolor[HTML]{FFFFE0}  0}                               & \multicolumn{ 1}{c|}{\cellcolor[HTML]{FFFFE0}  0}                                                                                        & \multicolumn{ 1}{c|}{\cellcolor[HTML]{FFFFE0}  1}                               & \multicolumn{ 1}{c|}{\cellcolor[HTML]{FFFFE0} 4}                               & \multicolumn{ 1}{c|}{\cellcolor[HTML]{FFFFE0}  0}                                                                                        & \multicolumn{ 1}{c|}{\cellcolor[HTML]{FFFFE0}  0}                               & {\cellcolor[HTML]{FFFFE0} 6}                               \\ \hline
 \multicolumn{ 1}{|l|}{\emph{\textbf{D:}} Challenging   Valleys}        & \multicolumn{ 1}{c|}{\cellcolor[HTML]{FFFFE0} 3}                                                                                                                & \multicolumn{ 1}{c|}{\cellcolor[HTML]{FFFFE0} 0}                               & \multicolumn{ 1}{c|}{\cellcolor[HTML]{FFFFE0} 0}                                                    & \multicolumn{ 1}{c|}{\cellcolor[HTML]{FFFFE0}  0}                                                                                                               & \multicolumn{ 1}{c|}{\cellcolor[HTML]{FFFFE0} 6}                               & \multicolumn{ 1}{c|}{\cellcolor[HTML]{FFFFE0}  0}                               & \multicolumn{ 1}{c|}{\cellcolor[HTML]{FFFFE0} 4}                                                                                        & \multicolumn{ 1}{c|}{\cellcolor[HTML]{FFFFE0} 2}                               & \multicolumn{ 1}{c|}{\cellcolor[HTML]{FFFFE0}  0}                               & \multicolumn{ 1}{c|}{\cellcolor[HTML]{FFFFE0}  0}                                                                                        & \multicolumn{ 1}{c|}{\cellcolor[HTML]{FFFFE0}  0}                               & \multicolumn{ 1}{c|}{\cellcolor[HTML]{FFFFE0}  0}                               & \multicolumn{ 1}{c|}{\cellcolor[HTML]{FFFFE0}  0}                                                                                        & \multicolumn{ 1}{c|}{\cellcolor[HTML]{FFFFE0} 2}                               & {\cellcolor[HTML]{FFFFE0} 10}                              \\ \hline
 \multicolumn{ 1}{|l|}{\emph{\textbf{E:}} Binary   Inversions}          & \multicolumn{ 1}{c|}{\cellcolor[HTML]{FFFFE0} 0}                                                                                                                & \multicolumn{ 1}{c|}{\cellcolor[HTML]{FFFFE0} 0}                               & \multicolumn{ 1}{c|}{\cellcolor[HTML]{FFFFE0} 0}                                                    & \multicolumn{ 1}{c|}{\cellcolor[HTML]{FFFFE0}  0}                                                                                                               & \multicolumn{ 1}{c|}{\cellcolor[HTML]{FFFFE0}  0}                               & \multicolumn{ 1}{c|}{\cellcolor[HTML]{FFFFE0}  0}                               & \multicolumn{ 1}{c|}{\cellcolor[HTML]{FFFFE0}  0}                                                                                        & \multicolumn{ 1}{c|}{\cellcolor[HTML]{FFFFE0} 5}                               & \multicolumn{ 1}{c|}{\cellcolor[HTML]{FFFFE0}  0}                               & \multicolumn{ 1}{c|}{\cellcolor[HTML]{FFFFE0}  0}                                                                                        & \multicolumn{ 1}{c|}{\cellcolor[HTML]{FFFFE0}  0}                               & \multicolumn{ 1}{c|}{\cellcolor[HTML]{FFFFE0} 6}                               & \multicolumn{ 1}{c|}{\cellcolor[HTML]{FFFFE0}  0}                                                                                        & \multicolumn{ 1}{c|}{\cellcolor[HTML]{FFFFE0} 4}                               & {\cellcolor[HTML]{FFFFE0} 4}                               \\ \hline
 \multicolumn{ 1}{|l|}{\emph{\textbf{F:}} Quests}                       & \multicolumn{ 1}{c|}{\cellcolor[HTML]{FFFFE0} 6}                                                                                                                & \multicolumn{ 1}{c|}{\cellcolor[HTML]{FFFFE0} 0}                               & \multicolumn{ 1}{c|}{\cellcolor[HTML]{FFFFE0} 0}                                                    & \multicolumn{ 1}{c|}{\cellcolor[HTML]{FFFFE0}  0}                                                                                                               & \multicolumn{ 1}{c|}{\cellcolor[HTML]{FFFFE0} 2}                               & \multicolumn{ 1}{c|}{\cellcolor[HTML]{FFFFE0}  0}                               & \multicolumn{ 1}{c|}{\cellcolor[HTML]{FFFFE0}  0}                                                                                        & \multicolumn{ 1}{c|}{\cellcolor[HTML]{FFFFE0} 8}                               & \multicolumn{ 1}{c|}{\cellcolor[HTML]{FFFFE0}  0}                               & \multicolumn{ 1}{c|}{\cellcolor[HTML]{FFFFE0}  0}                                                                                        & \multicolumn{ 1}{c|}{\cellcolor[HTML]{FFFFE0}  0}                               & \multicolumn{ 1}{c|}{\cellcolor[HTML]{FFFFE0} 2}                               & \multicolumn{ 1}{c|}{\cellcolor[HTML]{FFFFE0}  0}                                                                                        & \multicolumn{ 1}{c|}{\cellcolor[HTML]{FFFFE0}  0}                               & {\cellcolor[HTML]{FFFFE0} 8}                               \\ \hline
 \multicolumn{ 1}{|l|}{\emph{\textbf{G:}} SlavicG's   Favorite} & \multicolumn{ 1}{c|}{\cellcolor[HTML]{FFFFE0} 0}                                                                                                                & \multicolumn{ 1}{c|}{\cellcolor[HTML]{FFFFE0} 0}                               & \multicolumn{ 1}{c|}{\cellcolor[HTML]{FFFFE0} 0}                                                            & \multicolumn{ 1}{c|}{\cellcolor[HTML]{FFFFE0}  0}                                                                                                               & \multicolumn{ 1}{c|}{\cellcolor[HTML]{FFFFE0}  0}                               & \multicolumn{ 1}{c|}{\cellcolor[HTML]{FFFFE0}  0}                               & \multicolumn{ 1}{c|}{\cellcolor[HTML]{FFFFE0}  0}                                                                                        & \multicolumn{ 1}{c|}{\cellcolor[HTML]{FFFFE0} 8}                               & \multicolumn{ 1}{c|}{\cellcolor[HTML]{FFFFE0}  0}                               & \multicolumn{ 1}{c|}{\cellcolor[HTML]{FFFFE0}  0}                                                                                        & \multicolumn{ 1}{c|}{\cellcolor[HTML]{FFFFE0} 2}                               & \multicolumn{ 1}{c|}{\cellcolor[HTML]{FFFFE0} 5}                               & \multicolumn{ 1}{c|}{\cellcolor[HTML]{FFFFE0}  0}                                                                                        & \multicolumn{ 1}{c|}{\cellcolor[HTML]{FFFFE0}  0}                               & {\cellcolor[HTML]{FFFFE0} 5}                               \\ \hline
 \multicolumn{ 1}{|c|}{\emph{\textbf{Average}}}                      & \multicolumn{ 1}{c|}{\cellcolor[HTML]{FFC1C1} \textbf{4.1}}                                                                                                                & \multicolumn{ 1}{c|}{\cellcolor[HTML]{FFC1C1} \textbf{0.7}}                             & \multicolumn{ 1}{c|}{\cellcolor[HTML]{FFC1C1} \textbf{0.0}}                        & \multicolumn{ 1}{c|}{\cellcolor[HTML]{FFC1C1} \textbf{0.0}}                                                                                                             & \multicolumn{ 1}{c|}{\cellcolor[HTML]{FFC1C1} \textbf{2.9}}                             & \multicolumn{ 1}{c|}{\cellcolor[HTML]{FFC1C1} \textbf{0.0}}                             & \multicolumn{ 1}{c|}{\cellcolor[HTML]{FFC1C1} \textbf{0.6}}                                                                                      & \multicolumn{ 1}{c|}{\cellcolor[HTML]{FFC1C1} \textbf{4.4}}                             & \multicolumn{ 1}{c|}{\cellcolor[HTML]{FFC1C1} \textbf{0.0}}                             & \multicolumn{ 1}{c|}{\cellcolor[HTML]{FFC1C1} \textbf{0.0}}                                                                                      & \multicolumn{ 1}{c|}{\cellcolor[HTML]{FFC1C1} \textbf{0.7}}                             & \multicolumn{ 1}{c|}{\cellcolor[HTML]{FFC1C1} \textbf{3.0}}                             & \multicolumn{ 1}{c|}{\cellcolor[HTML]{FFC1C1} \textbf{0.0}}                                                                                      & \multicolumn{ 1}{c|}{\cellcolor[HTML]{FFC1C1} \textbf{1.1}}                             & \cellcolor[HTML]{FFC1C1} \textbf{7.0}                             \\ \hline
 \multicolumn{16}{|c|}{\emph{\textbf{Correct Program}}}                                                                                                                                                                                                                                                                                                                                                                                                                                                                                                                                                                                                                                                                                                                                                                                                                                                                                                                                                                                                                                                                                                                                                                                             \\ \hline
 \multicolumn{ 1}{|l|}{\emph{\textbf{A:}} Medium Number}                & \multicolumn{ 1}{c|}{\cellcolor[HTML]{FFFFE0} 0}                                                                                                                & \multicolumn{ 1}{c|}{\cellcolor[HTML]{FFFFE0} 0}                               & \multicolumn{ 1}{c|}{\cellcolor[HTML]{FFFFE0} 0}                                                    & \multicolumn{ 1}{c|}{\cellcolor[HTML]{FFFFE0}  0}                                                                                                               & \multicolumn{ 1}{c|}{\cellcolor[HTML]{FFFFE0}  0}                               & \multicolumn{ 1}{c|}{\cellcolor[HTML]{FFFFE0}  0}                               & \multicolumn{ 1}{c|}{\cellcolor[HTML]{FFFFE0}  0}                                                                                        & \multicolumn{ 1}{c|}{\cellcolor[HTML]{FFFFE0} 2}                               & \multicolumn{ 1}{c|}{\cellcolor[HTML]{FFFFE0}  0}                               & \multicolumn{ 1}{c|}{\cellcolor[HTML]{FFFFE0}  0}                                                                                        & \multicolumn{ 1}{c|}{\cellcolor[HTML]{FFFFE0} 3}                               & \multicolumn{ 1}{c|}{\cellcolor[HTML]{FFFFE0} 2}                               & \multicolumn{ 1}{c|}{\cellcolor[HTML]{FFFFE0}  0}                                                                                        & \multicolumn{ 1}{c|}{\cellcolor[HTML]{FFFFE0} 5}                               & {\cellcolor[HTML]{FFFFE0} 8}                               \\ \hline
 \multicolumn{ 1}{|l|}{\emph{\textbf{B:}} Atilla's   Favorite}  & \multicolumn{ 1}{c|}{\cellcolor[HTML]{FFFFE0} 0}                                                                                                                & \multicolumn{ 1}{c|}{\cellcolor[HTML]{FFFFE0} 0}                               & \multicolumn{ 1}{c|}{\cellcolor[HTML]{FFFFE0} 0}                                                            & \multicolumn{ 1}{c|}{\cellcolor[HTML]{FFFFE0}  0}                                                                                                               & \multicolumn{ 1}{c|}{\cellcolor[HTML]{FFFFE0}  0}                               & \multicolumn{ 1}{c|}{\cellcolor[HTML]{FFFFE0}  0}                               & \multicolumn{ 1}{c|}{\cellcolor[HTML]{FFFFE0}  0}                                                                                        & \multicolumn{ 1}{c|}{\cellcolor[HTML]{FFFFE0} 5}                               & \multicolumn{ 1}{c|}{\cellcolor[HTML]{FFFFE0}  0}                               & \multicolumn{ 1}{c|}{\cellcolor[HTML]{FFFFE0}  0}                                                                                        & \multicolumn{ 1}{c|}{\cellcolor[HTML]{FFFFE0} 5}                               & \multicolumn{ 1}{c|}{\cellcolor[HTML]{FFFFE0} 3}                               & \multicolumn{ 1}{c|}{\cellcolor[HTML]{FFFFE0}  0}                                                                                        & \multicolumn{ 1}{c|}{\cellcolor[HTML]{FFFFE0}  0}                               & {\cellcolor[HTML]{FFFFE0} 7}                               \\ \hline
 \multicolumn{ 1}{|l|}{\emph{\textbf{C:}} Advantage}                    & \multicolumn{ 1}{c|}{\cellcolor[HTML]{FFFFE0} 0}                                                                                                                & \multicolumn{ 1}{c|}{\cellcolor[HTML]{FFFFE0} 0}                               & \multicolumn{ 1}{c|}{\cellcolor[HTML]{FFFFE0} 0}                                                    & \multicolumn{ 1}{c|}{\cellcolor[HTML]{FFFFE0}  0}                                                                                                               & \multicolumn{ 1}{c|}{\cellcolor[HTML]{FFFFE0}  0}                               & \multicolumn{ 1}{c|}{\cellcolor[HTML]{FFFFE0}  0}                               & \multicolumn{ 1}{c|}{\cellcolor[HTML]{FFFFE0}  0}                                                                                        & \multicolumn{ 1}{c|}{\cellcolor[HTML]{FFFFE0} 7}                               & \multicolumn{ 1}{c|}{\cellcolor[HTML]{FFFFE0}  0}                               & \multicolumn{ 1}{c|}{\cellcolor[HTML]{FFFFE0}  0}                                                                                        & \multicolumn{ 1}{c|}{\cellcolor[HTML]{FFFFE0} 3}                               & \multicolumn{ 1}{c|}{\cellcolor[HTML]{FFFFE0} 4}                               & \multicolumn{ 1}{c|}{\cellcolor[HTML]{FFFFE0}  0}                                                                                        & \multicolumn{ 1}{c|}{\cellcolor[HTML]{FFFFE0}  0}                               & {\cellcolor[HTML]{FFFFE0} 6}                               \\ \hline
 \multicolumn{ 1}{|l|}{\emph{\textbf{D:}} Challenging Valleys}        & \multicolumn{ 1}{c|}{\cellcolor[HTML]{FFFFE0} 0}                                                                                                                & \multicolumn{ 1}{c|}{\cellcolor[HTML]{FFFFE0} 0}                               & \multicolumn{ 1}{c|}{\cellcolor[HTML]{FFFFE0} 0}                                                      & \multicolumn{ 1}{c|}{\cellcolor[HTML]{FFFFE0}  0}                                                                                                               & \multicolumn{ 1}{c|}{\cellcolor[HTML]{FFFFE0}  0}                               & \multicolumn{ 1}{c|}{\cellcolor[HTML]{FFFFE0}  0}                               & \multicolumn{ 1}{c|}{\cellcolor[HTML]{FFFFE0} 3}                                                                                        & \multicolumn{ 1}{c|}{\cellcolor[HTML]{FFFFE0}  0}                               & \multicolumn{ 1}{c|}{\cellcolor[HTML]{FFFFE0}  0}                               & \multicolumn{ 1}{c|}{\cellcolor[HTML]{FFFFE0}  0}                                                                                        & \multicolumn{ 1}{c|}{\cellcolor[HTML]{FFFFE0}  1}                               & \multicolumn{ 1}{c|}{\cellcolor[HTML]{FFFFE0}  0}                               & \multicolumn{ 1}{c|}{\cellcolor[HTML]{FFFFE0}  0}                                                                                        & \multicolumn{ 1}{c|}{\cellcolor[HTML]{FFFFE0} 8}                               & {\cellcolor[HTML]{FFFFE0} 10}                              \\ \hline
 \multicolumn{ 1}{|l|}{\emph{\textbf{E:}} Binary   Inversions}          & \multicolumn{ 1}{c|}{\cellcolor[HTML]{FFFFE0} 0}                                                                                                                & \multicolumn{ 1}{c|}{\cellcolor[HTML]{FFFFE0} 0}                               & \multicolumn{ 1}{c|}{\cellcolor[HTML]{FFFFE0} 0}                                                    & \multicolumn{ 1}{c|}{\cellcolor[HTML]{FFFFE0}  0}                                                                                                               & \multicolumn{ 1}{c|}{\cellcolor[HTML]{FFFFE0}  0}                               & \multicolumn{ 1}{c|}{\cellcolor[HTML]{FFFFE0}  0}                               & \multicolumn{ 1}{c|}{\cellcolor[HTML]{FFFFE0}  0}                                                                                        & \multicolumn{ 1}{c|}{\cellcolor[HTML]{FFFFE0} 6}                               & \multicolumn{ 1}{c|}{\cellcolor[HTML]{FFFFE0}  0}                               & \multicolumn{ 1}{c|}{\cellcolor[HTML]{FFFFE0}  0}                                                                                        & \multicolumn{ 1}{c|}{\cellcolor[HTML]{FFFFE0}  0}                               & \multicolumn{ 1}{c|}{\cellcolor[HTML]{FFFFE0} 2}                               & \multicolumn{ 1}{c|}{\cellcolor[HTML]{FFFFE0}  0}                                                                                        & \multicolumn{ 1}{c|}{\cellcolor[HTML]{FFFFE0} 4}                               & {\cellcolor[HTML]{FFFFE0} 8}                               \\ \hline
 \multicolumn{ 1}{|l|}{\emph{\textbf{F:}} Quests}                       & \multicolumn{ 1}{c|}{\cellcolor[HTML]{FFFFE0} 0}                                                                                                                & \multicolumn{ 1}{c|}{\cellcolor[HTML]{FFFFE0} 0}                               & \multicolumn{ 1}{c|}{\cellcolor[HTML]{FFFFE0} 0}                                                    & \multicolumn{ 1}{c|}{\cellcolor[HTML]{FFFFE0}  0}                                                                                                               & \multicolumn{ 1}{c|}{\cellcolor[HTML]{FFFFE0}  0}                               & \multicolumn{ 1}{c|}{\cellcolor[HTML]{FFFFE0}  0}                               & \multicolumn{ 1}{c|}{\cellcolor[HTML]{FFFFE0}  0}                                                                                        & \multicolumn{ 1}{c|}{\cellcolor[HTML]{FFFFE0} 7}                               & \multicolumn{ 1}{c|}{\cellcolor[HTML]{FFFFE0}  0}                               & \multicolumn{ 1}{c|}{\cellcolor[HTML]{FFFFE0}  0}                                                                                        & \multicolumn{ 1}{c|}{\cellcolor[HTML]{FFFFE0}  1}                               & \multicolumn{ 1}{c|}{\cellcolor[HTML]{FFFFE0}  1}                               & \multicolumn{ 1}{c|}{\cellcolor[HTML]{FFFFE0}  0}                                                                                        & \multicolumn{ 1}{c|}{\cellcolor[HTML]{FFFFE0} 2}                               & {\cellcolor[HTML]{FFFFE0} 9}                               \\ \hline
 \multicolumn{ 1}{|l|}{\emph{\textbf{G:}} SlavicG's Favorite}   & \multicolumn{ 1}{c|}{\cellcolor[HTML]{FFFFE0} 0}                                                                                                                & \multicolumn{ 1}{c|}{\cellcolor[HTML]{FFFFE0} 0}                               & \multicolumn{ 1}{c|}{\cellcolor[HTML]{FFFFE0} 0}                                                            & \multicolumn{ 1}{c|}{\cellcolor[HTML]{FFFFE0}  0}                                                                                                               & \multicolumn{ 1}{c|}{\cellcolor[HTML]{FFFFE0}  0}                               & \multicolumn{ 1}{c|}{\cellcolor[HTML]{FFFFE0}  0}                               & \multicolumn{ 1}{c|}{\cellcolor[HTML]{FFFFE0}  0}                                                                                        & \multicolumn{ 1}{c|}{\cellcolor[HTML]{FFFFE0} 3}                               & \multicolumn{ 1}{c|}{\cellcolor[HTML]{FFFFE0}  0}                               & \multicolumn{ 1}{c|}{\cellcolor[HTML]{FFFFE0}  0}                                                                                        & \multicolumn{ 1}{c|}{\cellcolor[HTML]{FFFFE0}  1}                               & \multicolumn{ 1}{c|}{\cellcolor[HTML]{FFFFE0} 4}                               & \multicolumn{ 1}{c|}{\cellcolor[HTML]{FFFFE0}  0}                                                                                        & \multicolumn{ 1}{c|}{\cellcolor[HTML]{FFFFE0} 5}                               & {\cellcolor[HTML]{FFFFE0} 6}                               \\ \hline
 \multicolumn{ 1}{|c|}{\emph{\textbf{Average}}}                      & \multicolumn{ 1}{c|}{\cellcolor[HTML]{FFC1C1} \textbf{0.0}}                                                                                                              & \multicolumn{ 1}{c|}{\cellcolor[HTML]{FFC1C1} \textbf{0.0}}                             & \multicolumn{ 1}{c|}{\cellcolor[HTML]{FFC1C1} \textbf{0.0}}                             & \multicolumn{ 1}{c|}{\cellcolor[HTML]{FFC1C1} \textbf{0.0}}                                                                                                             & \multicolumn{ 1}{c|}{\cellcolor[HTML]{FFC1C1} \textbf{0.0}}                             & \multicolumn{ 1}{c|}{\cellcolor[HTML]{FFC1C1} \textbf{0.0}}                             & \multicolumn{ 1}{c|}{\cellcolor[HTML]{FFC1C1} \textbf{0.4}}                                                                                      & \multicolumn{ 1}{c|}{\cellcolor[HTML]{FFC1C1} \textbf{4.3}}                             & \multicolumn{ 1}{c|}{\cellcolor[HTML]{FFC1C1} \textbf{0.0}}                             & \multicolumn{ 1}{c|}{\cellcolor[HTML]{FFC1C1} \textbf{0.0}}                                                                                      & \multicolumn{ 1}{c|}{\cellcolor[HTML]{FFC1C1} \textbf{2.0}}                             & \multicolumn{ 1}{c|}{\cellcolor[HTML]{FFC1C1} \textbf{2.3}}                             & \multicolumn{ 1}{c|}{\cellcolor[HTML]{FFC1C1} \textbf{0.0}}                                                                                      & \multicolumn{ 1}{c|}{\cellcolor[HTML]{FFC1C1} \textbf{3.4}}                             & \cellcolor[HTML]{FFC1C1} \textbf{7.7}                             \\  \bottomrule
 \multicolumn{16}{l}{$^\dag$\emph{Each cell of the table shows the number of \ft, \fc, \ff, \ft and \iat found by \xxx or the baselines for each subject of \codeforces.}}\\
 \multicolumn{16}{l}{$^\dag$\emph{For each subject, \xxx and the baselines are run ten times.}} \\
 \multicolumn{16}{l}{$^\dag$\emph{\update{The total number of test cases found by \xxx for a subject can be less than ten, because \xxx may not return any test case in an execution.}}}\\
 %\multicolumn{16}{l}{\ying{\ft, \fc, \ff, \pt and \iat are not used in this table? Please check it.}}\\
 %\multicolumn{16}{l}{$^\dag$\emph{Adjusted average only considers five out of seven subjects, see S\ref{sec:eva:rq4} for details.}}
 
 % \vspace{-7mm}
 \end{tabular}
 \egroup
 \label{tab:rq4}
 \vspace{-5mm}
 \end{table*}

\subsection{RQ4: Finding \ft for \codeforces}\label{sec:eva:rq4}

\subsubsection{Experiment setup}

To address the validity threat due to data leakage discussed in \S~\ref{sec:eva:dataleakage}, \update{we conduct an evaluation on \codeforces programs released after the cutoff date of ChatGPT's training dataset}.  %\codeforces~\cite{codeforces} is a programming contest portal that stores programs submitted by previous participants. 
%To examine the performance of \xxx and baselines on the program subjects that have not been seen by ChatGPT, all selected subjects from \codeforces were created after the cut-off date of ChatGPT's training dataset. %These programs usually include both buggy programs and correct programs of a contest problem, because a participant can make several submissions. \codeforces provides test cases to verify the correctness of a program. Hence, \codeforces programs can form a benchmark similar to \quixbugs. 
%\scc{To be moved to RQ4} We also select evaluation subjects from \codeforces~\cite{codeforces}, a popular programming contest portal because \codeforces is an official benchmark for ChatGPT~\cite{gpt4}. 

Specifically, we selected programs from a contest for programming beginners (named \textit{Codeforces Round 835}, held on 21, November 2022). \textit{Codeforces Round 835} is the educational contest most recent to the date that ChatGPT was launched (i.e., 30, November 2022). %We start experimenting ChatGPT after it is launched, so the most recent contest to the date of launch is chosen.
%We selected this contest because it is available right after 
%\scc{What is the benefit of selecting subjects recent to the date that ChatGPT was launched?}\tszon{There is no benefit. We may justify that we start experimenting ChatGPT after it is launched, so the most recent contest to the date of launch is chosen. Currently this is the best reason I can give, though may not be too convinicing.} \scc{Is it the most recent one when we started our study using gpt-turbo? People will naturally choose the more recent ones, but why we didn't do so?} \tszon{our pilot study includes the most recent subjects and subjects from this contest, and found that \xxx performs the best on subjects from this contest...let me think of a more convincing reason.} 
The contest contains seven programming problems of different difficulties. For each programming problem, we \update{choose} a buggy version and a correct version based on two criteria. First, the correct program has to be the bug-fixed version of the buggy program, so that we can compare \xxx's success rate and false positive rate fairly. Second, both programs are implemented in Python. If there is more than one candidate satisfying both criteria, we randomly choose one. %We randomly selected a subject that fulfills these two criteria. 
% \update{Since the selected programs have different complexity levels and solve less-known problems, these programs are representative of real-world programs.}

We use the two baselines as adopted for RQ1 (\S\ref{sec:eva:rq1}): \gptbaseline and \textsc{Pynguin}. We apply \xxx and the baselines to each program 10 times and record the number of \ftests found in each category, following the experiment setup in RQ1. We adopt the \textit{success rate} and \textit{accuracy} defined in RQ1 as the evaluation metric of this RQ.

\begin{figure}[t]
\centering
%\vspace{-2mm}
\includegraphics[width=0.55\linewidth]{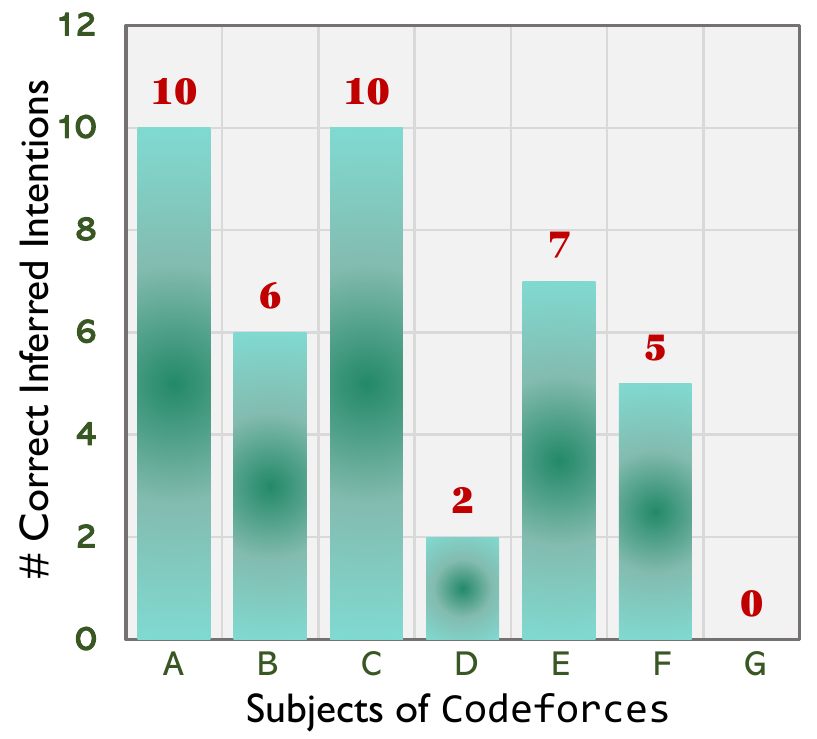}
\vspace{-2mm}
\caption{ChatGPT's effectiveness in inferring intentions for \update{buggy programs of \codeforces}.}
\vspace{-6mm}
\label{fig:eva:codeforce}
\end{figure}

\subsubsection{Results and findings}\label{sec:eva:rq4:findings}

Table~\ref{tab:rq4} shows that for buggy programs of \codeforces, \xxx's success rate (\xxxcodeforcesbuggyftoriginal) outperforms the best baseline (\gptcodeforcesbuggyftoriginal) by \xxxcodeforcesbuggyftvsbaselinesoriginal. In addition, for both buggy and correct programs, \xxx returns incorrect \ftests for only one subject ({\mycode Challenging Valleys}), while \gptbaseline returns incorrect \ftests for all seven subjects. Besides, \textsc{Pynguin} finds no correct \ftests in any executions. 

%\scc{Reviewers could ask if \textsc{Pynguin} can find passing tests that contain failure-inducing test inputs. Need to check the passing tests by \textsc{Pynguin} for subjects E and G may contain failure-inducing test inputs. The same case for Quixbugs. If \textsc{Pynguin} can find failure-inducing test inputs, we should consider generating test inputs using Pynguin. Indeed, our original idea is to use Evosuite to generate test inputs for Java subjects. It is fine if the generated test cases contain failure-inducing test inputs. It has been released for over one year. I don't think such an argument can convince reviewers not to use Pynguin. We need to understand that such a question can be easily raised by a reviewer. I meant if \textsc{Pynguin} turns out that it cannot find any failure-inducing test inputs in the generated passing test cases, it is easy to defend. Also check the Quixbugs ones where we argue that the root cause is the inability to reach the buggy branch.}\tszon{Got it. start working on it and will put a conclusion in this paragraph. \textsc{Pynguin} often generates less than 10 test cases before termination (whether reached 100\% coverage or not), and we cannot configure the stopping condition. I believe it is an implementation issue with Pynguin. It is essentially a new tool released in 2022. I believe using Evosuite will be much better. No it won't. just that it is my observation. I will think of an argument. I can image that, let me find a convincing explanation. Ok I will check. Got it.}

Note that \xxx's success rate on \codeforces is lower than that on \quixbugs. \update{Table~\ref{tab:rq1accuracy} also shows that \xxx's accuracy is notably lower on \codeforces than that on \quixbugs}. We observe that such discrepancies can be caused by differences in complexities/difficulties between programs of \codeforces and \quixbugs. 
%Intuitively, a \ft is more difficult to find for more complicated programs. %Since the selected \codeforces subjects have different difficulties, \xxx's overall success rate is inevitably dragged down by \xxx's performance on those complicated subjects.

Specifically, Table~\ref{tab:rq4} shows the seven selected \codeforces programs. \update{In fact, \codeforces has assigned a label (a letter from \emph{A} to \emph{G}) to each program: the label represents the complexity level of a program. We also present these labels in Table~\ref{tab:rq4}.} Particularly, the program labeled with \emph{A} is the simplest and the program labeled with \emph{G} is the most difficult. Essentially, \xxx performs the best (i.e., finds ten \ft) for \emph{A}, and its performance gradually decreases for programs with increasing difficulties. \update{We also find that \xxx's performance in inferring intentions decreases for programs with increasing difficulties (Figure~\ref{fig:eva:codeforce}). }

We further analyze \codeforces' programs to understand how program complexity may affect \xxx's performance.
%Actually, subjects between \emph{P4} to \emph{P7} are more complex than those of \quixbugs. 
Programs with label \emph{D} to \emph{G} are rated with a difficulty level of at least 1000 in \codeforces. According to studies~\cite{ebtekar2021elo,codeforceslevel} of \codeforces' rating scheme, programs that are rated 1000 or above focus on applying advanced data structures or algorithmic techniques (e.g., divide-and-conquer, dynamic programming) for solving complex problems. 

In comparison, programs with label \emph{A} to \emph{C} are rated with a difficulty level of 800, according to the studies~\cite{ebtekar2021elo,codeforceslevel}, they focus on implementing common algorithms (e.g., \emph{A} implements the computation of medium number). Hence, \emph{A} to \emph{C} have similar complexities with \quixbugs' programs, because \quixbugs also focuses on implementing common algorithms adopted by real-world programs~\cite{lin2017quixbugs}, such as {\mycode quicksort}, {\mycode detect-cycle}, {\mycode reverse-linked-list}. %Among all programs of \quixbugs, 

%Only one exception exists ({\mycode lcs-length} implements a dynamic algorithm, as discussed in \S\ref{fig:eva:intention}).%Among all \quixbugs subjects {\mycode lcs-length} (discussed in \S\ref{fig:eva:intention}) is the only exception. 

%Hence, \quixbugs's complexity is similar to \emph{\textbf{P1}} to \emph{\textbf{P3}} (which are rated below 1000) 

% Hence

% which implies the subjects solve a logical puzzle (e.g., optimizing a brute-force algorithm). Hence, the program semantics can be more abstract, and . On the other hand, \emph{\textbf{P1}} to \emph{\textbf{P4}} are tagged with \textit{implementation}, which means 
We find that \xxx's success rate over \emph{A} to \emph{C} (\xxxcodeforcesbuggyftimproved) is not significantly less than its success rate over the \quixbugs programs (\xxxft). The results suggest that \xxx is able to find \ftests for \codeforces programs with similar complexity as \quixbugs programs. \update{Similarly, \xxx's accuracy over \emph{A} to \emph{C} is 100\% because no incorrect \ftest has been found. It is comparable with that on \quixbugs programs (over 94.6\%, see Table~\ref{tab:rq1breakdown})}. 

However, \xxx may not work well on complex programs. An interesting future research direction is to explore a reduction methodology, allowing the use of \xxx on the simpler code snippets reduced from a complex program. %Given that programs of comparable complexity to \quixbugs programs are pervasive. 
Experimental results suggest that \xxx is effective to find for those programs the failure-inducing test cases, that the state-of-the-art test generation technique and a strawman approach of using ChatGPT are unlikely able to find.

\begin{tcolorbox}[boxrule=1pt,boxsep=1pt,left=2pt,right=2pt,top=2pt,bottom=2pt]
% \small
 \textcolor{mygreen}{\faIcon{key}} \noindent\emph{\textbf{Finding 5:}} \emph{For selected \codeforces programs that have similar complexity with \quixbugs programs, \xxx's success rate on these \codeforces programs (\xxxcodeforcesbuggyftimproved) is comparable to its success rate on \quixbugs programs (\xxxft).
 }
	
 \textcolor{red}{\faIcon{user-edit}} \noindent\emph{\textbf{Implication:}} \emph{
 \xxx performs comparably on \quixbugs and \codeforces programs with similar complexity. %\tszon{sounds repetitive, shall we mention data leakage?}
 }
\end{tcolorbox} 

\subsection{Threat to Validity}

Our evaluation can be subject to several validity threats.

\subsubsection{Data leakage}\label{sec:eva:dataleakage}

``Data leakage'' refers to the problem that an evaluation is conducted on a dataset that has been included in ChatGPT's training dataset. In this case, overfitting may occur and cause evaluation results to be biased. For instance, \xxx's evaluation is conducted on {\mycode QuixBugs}, a public benchmark released in 2017~\cite{lin2017quixbugs}. It is possible that the benchmark has been used to train ChatGPT whose training cutoff date is September 2021. If so, the performance of both \gptbaseline and \xxx may be overestimated.

Given the low success rate (\gptft) of \gptbaseline in finding \ftests, the data leakage threat with ChatGPT is likely to be insignificant. %However, we argue that even {\mycode Quixbugs} has been included in ChatGPT's training dataset, the overestimation is unlikely to be significant; otherwise \gptbaseline's success rate in finding \ftests would not be that unsatisfactory (\gptft). %Suppose {\mycode Quixbugs} has been included in ChatGPT's training dataset, ChatGPT's actual success rate will be even lower. Hence, our evaluation result provides a good estimate of ChatGPT's strengths and limitations in finding \ftests in diverse real-world scenarios.
In addition, we mitigated the threat with additional experiments on \codeforces programs, which were created after the training cut-off date of ChatGPT. The performance of \xxx on the \codeforces programs is comparable with that on the \quixbugs programs. %, which implies \xxx's effectiveness is unlikely a result of data leakage.

% \subsubsection{Limited number of evaluation subjects}

% We only conducted experiment on forty programs of Quixbugs. Hence, the scale of our evaluation may not fully reveals \xxx's strength and limitations. However, Quixbugs consists of diverse basic algorithms which are widely adopted in real-world software~\cite{lin2017quixbugs,xia2023conversational,sobania2023analysis}. This result shows that our evaluation result can be a good estimator of wide-range of real world software.

\subsubsection{Generalizability}\label{sec:threat:simple}

Evaluation is made only on \quixbugs and \codeforces programs. These programs are simple programs that have less than one hundred lines of code. Hence, there would be a concern that our evaluation result on \xxx may not be generalized to large real-world software. Nonetheless, as pointed out by Weimer et al.~\cite{weimer2009automatically}, large real-world software is often made up of small programs. %, and these small programs are common across different software. 
% For instance, array manipulation functions implemented in Quixbugs' programs). 
Hence, our evaluation result reveals \xxx's potential of being applied to large real-world software. In addition, \xxx's enhancement over baselines is a crucial step towards finding \ftests for large real-world software.

% Finally, this paper chose ChatGPT as the LLM to study, because ChatGPT is the only LLM that offers API interface and has unlimited access. Studies related on other LLMs (e.g., GPT-4) are future work

%This paper studies and adopts a specific LLM \textit{gpt-turbo}. 
Another possible validity threat is that our evaluation result based on \textit{gpt-3.5-turbo-0301} may not be generalized to other LLMs such as GPT-4. Nevertheless, we choose \textit{gpt-3.5-turbo-0301} because it is the only LLM providing an API interface that is not subject to limited access. Besides, the technical report~\cite{gpt4} released by OpenAI shows that GPT-4, the state-of-the-art LLM, does not outperform ChatGPT (the backbone model of \textit{gpt-3.5-turbo-0301}) for code-related tasks.
 Hence, our results can provide a useful reference for future related studies. 

\subsubsection{Reproducibility}\label{sec:threat:reproduce}

\update{Our evaluation is conducted using \textit{gpt-3.5-turbo-0301}. One possible threat is that the evaluation results presented may not be reproducible after the model has been deprecated in the future. Nevertheless, since \xxx targets at addressing the fundamental limitations of LLMs and conventional approaches (e.g., Pynguin) in finding \ftests (\S\ref{sec:methdology}), \xxx's enhancement over existing approaches is unlikely subject to specific LLMs. Besides, as discussed in \S\ref{sec:threat:simple}, more advanced LLMs (e.g., GPT-4) do not necessarily outperform \textit{gpt-3.5-turbo-0301}. Hence, the results and findings in our paper can still provide useful references even \textit{gpt-3.5-turbo-0301} is deprecated.}

%Since GPT-4 is already one of the most advanced LLMs, our evaluation result essentially reveals the true strength and limitations that modern LLMs
%The technical report also implies that larger or more advanced LLM can unlikely achieve better performances. Hence,  

% \subsubsection{Generalizability}\label{sec:threat:gpt}

% hence \xxx has inferred ten intentions for each programs. We collect these intentions and manually investigate whether the intention is correct. Given an intention inferred by \xxx, two PhD students individually determine whether the intention is correct, and reach a consensus with the moderation of the students' supervisor. 

% To exhibit \xxx's usefulness, we repair the buggy programs with failing test identified \xxx. 

% We compare \xxx with two baselines: directly

% Developers can repair a buggy program with the identified failing test.

\section{Discussion and future work}\label{sec:discussion}

% \subsection{Future work}
Compared to baselines, \xxx has a notably higher success rate in finding \ftests (\S\ref{sec:eva:rq1}). We analyze the few cases where \xxx cannot find \ftests and find that the bugs are often located at a buggy branch which can be reached only by specific test input values (see \S\ref{sec:eva:rq1}). %we identified the main reason is that \xxx cannot find a test input that can trigger a PUT's failure. Moreover, we found that for these PUTs, test inputs generated by \xxx often cannot reach a buggy branch of a PUT.

A possible enhancement is to augment the prompting with test coverage guidance. For instance, \xxx can conversationally inform ChatGPT about uncovered statements, and request ChatGPT to generate test inputs to cover those statements. In doing so, ChatGPT can have a higher probability of suggesting failure-inducing test inputs. 

\update{Besides augmenting the prompting, \xxx can be readily adapted to Coverage-based Test Generation by 1) leveraging state-of-the-art coverage-based test generation techniques, which generate a set of diverse test cases, and 2) constructing the corresponding test oracles for these test cases using the reference versions. }

Apart from test input generation, in this paper, \xxx focuses on finding \ftests for relatively simple programs. A future study is to decompose a large program into small programs, and then deduce \ftests from the \ftests found for these small programs. %so that \xxx first finds \ftests for small programs, and then combines those \ftests into complete \ftests for the large program. 

%This paper focuses on leveraging ChatGPT to find a failing test (\S\ref{sec:intro}). 
% Finally, this paper chose ChatGPT as the LLM to study, because ChatGPT is the only LLM that offers API interface and has unlimited access. Studies related on other LLMs (e.g., GPT-4) are future work.

\section{Related Work}\label{sec:relatedwork}

% \subsection{Finding fault triggering tests}

Related works mainly fall into the three categories below.

\vspace{-2mm}
\subsection{Finding failure-inducing test inputs}
% \tszon{This subsection is not ready for proofreading.}

% Traditional automated test generation

\update{There are many pieces of works that study the problem of test input generation (e.g., coverage-guided test input generation~\cite{lukasczyk2022pynguin,fraser2011evosuite}, symbolic execution~\cite{baldoni2018survey} etc). However, not all of them focus on finding failure-inducing test inputs (e.g., Pynguin~\cite{lukasczyk2022pynguin} focuses on generating test inputs that achieve high code coverage). To the best of our knowledge, the most closely-related works would be those target the problem of failing test reproduction~\cite{23,24,25}, failing test augmentation~\cite{zhang2022improving,27}, or fuzzing~\cite{manes2019art,saha2023rare,liang2019deepfuzzer}. Failing test reproduction studies the problem of generating failure-inducing test inputs from bug reports. Failing test augmentation studies the problem of generating failure-inducing test inputs based on existing failure-inducing test cases. Failing test augmentation is useful for fault localization. Essentially these works are orthogonal to \xxx, because these works assume a PUT is buggy and additional information (e.g., bug reports or failing tests) is available. In contrast, \xxx is provided with a PUT only (without having knowledge of whether the PUT is buggy or not).}

\update{Regarding fuzzing, fuzzing techniques focus on inducing specific types of failures (e.g., crash, security vulnerabilities) that the corresponding test oracles have been pre-defined~\cite{manes2019art}. In comparison, \xxx is not restricted to the detection of specific failure types. Furthermore, it does not require test oracles to be pre-defined.}

\vspace{-2mm}
\subsection{Addressing Test Oracle problem}

% The correct behavior can be inferred from execution 
\update{Automated test oracle construction is a longstanding challenge. Several paradigms have been proposed to address this challenge, such as metamorphic testing~\cite{chen2020metamorphic,tian2021extent,cao2022semmt,segura2016survey} and differential testing~\cite{mckeeman1998differential,evans2007differential,chen2016coverage}. However, these paradigms often have limited application scenarios~\cite{ibrahimzada2022perfect}. 
% Jahangirova et al.~\cite{jahangirova2016test, jahangirova2017oracle} develop a technique for assessing and improving test oracles by reducing the incidence of both false positives and false negatives. They prove that their approach can always result in an increase in the mutual information between the actual and perfect oracles.
% Afterwards, they propose a human-in-the-loop approach~\cite{jahangirova2019empirical} for oracle improvement and analyse whether the proposed oracle improvement process is helping developers to create better oracles. 
A recent technique SEER~\cite{ibrahimzada2022perfect} aims to address this limitation by proposing a framework to train Deep Learning models for inferring a PUT's behaviors. SEER and \xxx are orthogonal because \xxx focuses on leveraging existing commercial LLMs (e.g., ChatGPT) to address the test oracle problem, instead of training new models (which could be cost-prohibitive for LLMs).}

%However, as shown in SEER's paper, a developer needs to train SEER with code snippets closely related to a PUT, otherwise SEER's performance can have a significant drop. Moreover, such a training dataset may not be available in practice. In comparison, \xxx is integrated with ChatGPT, which has already been trained with diverse real-world software and has a much stronger inference performance than a standard Transformer model trained~\cite{zhou2023comprehensive}.

\vspace{-2mm}
\subsection{Studies of ChatGPT and other LLMs}

After ChatGPT has been released, several studies are proposed to study its effectiveness in tackling diverse software engineering problems. Dominik, et al.~\cite{sobania2023analysis} studies ChatGPT's effectiveness in generating patches. We adopt the prompting convention proposed by this work as the baseline of \xxx in our evaluation (\S\ref{sec:rq3:finding}). 
%\ying{The reference is broken.} 
There are also other works studying ChatGPT's (or other LLMs') capabilities for various software engineering tasks, including bug repair~\cite{xia2023keep, sobania2023analysis}, fuzzing~\cite{deng2023large}, code generation~\cite{tian2023chatgpt,vaithilingam2022expectation}, code summarization~\cite{ahmed2022few}, software testing education~\cite{jalil2023chatgpt}, and vulnerability detection~\cite{pearce2021can,xia2022practical}. To the best of our knowledge, no existing works study ChatGPT's (or other LLMs') capability in finding \ftests. This paper fills the research gap.
% Recent emergence of LLM 

% Previous techniques which construct a test oracle can be divided into three types: 

% Conventionally, the test oracle problem is addressed by relying on implicit oracle (e.g., hang, crash), or techniques such as metamorphic testing or differential testing. However, all these solutions are not applicable to general bugs or software.

% With recent advancement of deep learning, many recent works leverage deep learning to tackle the test oracle problem. 
\section{Conclusion}

In this paper, we propose \xxx, the first paradigm for finding \ftests using ChatGPT. The insight is that the program intention inferred by ChatGPT is insensitive to nuances in code. It allows ChatGPT to infer correct intention from a buggy program. With this insight, \xxx finds \ftests into three steps: program intention inference, program generation, and differential testing. %y doing so, \xxx bypasses the cons caused by ChatGPT's insensitivity to nuances, while fully enjoying its cons (i.e., ChatGPT is strong at inferring program intention, a crucial step in finding a \ftest). 
Our evaluation result shows that \xxx significantly outperforms state-of-the-art baselines in finding \ftests.

% In this paper, we proposed the first automated API-device
% correlation learning approach, PIVOT, to facilitate FIC issue
% detection. To effectively identify valid API-device correlations,
% PIVOT performs inter-procedural static analysis to extract APIdevice correlations and leverages a novel ranking strategy to
% prioritize them. The evaluation results show that our ranking
% strategy can effectively identify valid API-device correlations
% and significantly outperform an existing technique. Based on
% the learned API-device correlations, we built an archive of
% FIC issues and further conducted a case study to show the
% usefulness of API-device correlations. Our experiment results
% and other data are published at our project website [20].
% Currently, PIVOT requires human efforts to validate the
% learned API-device correlations. In future, we plan to study
% how to automate the validation process by combining program
% analysis and test synthesis techniques.

\vspace{-1mm}
\section{Data Availability}\label{sec:repro}

  We provide a reproduction package at \textbf{\url{https://differential-prompting.github.io/}} to facilitate future research. The package includes (1) a dataset containing 470 intentions inferred by \xxx, 940 reference versions, and all \ftests found by \xxx and the baselines (2) an available tool \xxx, and (3) seven \codeforces programs.
\section*{Acknowledgements}
\update{We would like to thank the anonymous reviewers for their comments and suggestions. We would also like to thank DL library developers for analyzing our reported issues. 
This work is supported by the National Science Foundation of China (Grant No. 61932021, 62141210), the Hong Kong Research Grant Council/General Research Fund (Grant No. 16205722), the Hong Kong Research Grant Council/Research Impact Fund (Grant No. R5034-18), the Fundamental Research Funds for the Central Universities (Grant No. N2217005), and Open Fund of State Key Lab. for Novel Software Technology, Nanjing University (KFKT2021B01).}

\balance
%\bibliographystyle{ACM-Reference-Format}
%\bibliography{sigproc}
%\bibliography{sigproc}
% This next section command marks the start of
% Appendix B, and does not continue the present hierarchy
%\newpage

%\balance
\bibliographystyle{IEEEtran}
\bibliography{main}

\begin{thebibliography}{10}
\providecommand{\url}[1]{#1}
\csname url@samestyle\endcsname
\providecommand{\newblock}{\relax}
\providecommand{\bibinfo}[2]{#2}
\providecommand{\BIBentrySTDinterwordspacing}{\spaceskip=0pt\relax}
\providecommand{\BIBentryALTinterwordstretchfactor}{4}
\providecommand{\BIBentryALTinterwordspacing}{\spaceskip=\fontdimen2\font plus
\BIBentryALTinterwordstretchfactor\fontdimen3\font minus
  \fontdimen4\font\relax}
\providecommand{\BIBforeignlanguage}[2]{{%
\expandafter\ifx\csname l@#1\endcsname\relax
\typeout{** WARNING: IEEEtran.bst: No hyphenation pattern has been}%
\typeout{** loaded for the language `#1'. Using the pattern for}%
\typeout{** the default language instead.}%
\else
\language=\csname l@#1\endcsname
\fi
#2}}
\providecommand{\BIBdecl}{\relax}
\BIBdecl

\bibitem{fraser2015does}
G.~Fraser, M.~Staats, P.~McMinn, A.~Arcuri, and F.~Padberg, ``Does automated
  unit test generation really help software testers? a controlled empirical
  study,'' \emph{ACM Transactions on Software Engineering and Methodology
  (TOSEM)}, vol.~24, no.~4, pp. 1--49, 2015.

\bibitem{ibrahimzada2022perfect}
A.~R. Ibrahimzada, Y.~Varli, D.~Tekinoglu, and R.~Jabbarvand, ``Perfect is the
  enemy of test oracle,'' in \emph{Proceedings of the 30th ACM Joint European
  Software Engineering Conference and Symposium on the Foundations of Software
  Engineering}, 2022, pp. 70--81.

\bibitem{sobania2023analysis}
D.~Sobania, M.~Briesch, C.~Hanna, and J.~Petke, ``An analysis of the automatic
  bug fixing performance of chatgpt,'' \emph{arXiv preprint arXiv:2301.08653},
  2023.

\bibitem{xia2023conversational}
C.~S. Xia and L.~Zhang, ``Conversational automated program repair,''
  \emph{arXiv preprint arXiv:2301.13246}, 2023.

\bibitem{chatgpt}
``Chatgpt: Optimizing language models for dialogue,''
  \url{https://openai.com/blog/chatgpt/}, 2023, accessed: 2023-04-01.

\bibitem{lin2017quixbugs}
D.~Lin, J.~Koppel, A.~Chen, and A.~Solar-Lezama, ``Quix{B}ugs: A multi-lingual
  program repair benchmark set based on the quixey challenge,'' in
  \emph{Proceedings Companion of the 2017 ACM SIGPLAN international conference
  on systems, programming, languages, and applications: software for humanity},
  2017, pp. 55--56.

\bibitem{zhang2020semantics}
Z.~Zhang, Y.~Wu, H.~Zhao, Z.~Li, S.~Zhang, X.~Zhou, and X.~Zhou,
  ``Semantics-aware bert for language understanding,'' in \emph{Proceedings of
  the AAAI Conference on Artificial Intelligence}, vol.~34, no.~05, 2020, pp.
  9628--9635.

\bibitem{zhang2021fast}
J.~Zhang, W.-C. Chang, H.-F. Yu, and I.~Dhillon, ``Fast multi-resolution
  transformer fine-tuning for extreme multi-label text classification,''
  \emph{Advances in Neural Information Processing Systems}, vol.~34, pp.
  7267--7280, 2021.

\bibitem{jiang2021lightxml}
T.~Jiang, D.~Wang, L.~Sun, H.~Yang, Z.~Zhao, and F.~Zhuang, ``Lightxml:
  Transformer with dynamic negative sampling for high-performance extreme
  multi-label text classification,'' in \emph{Proceedings of the AAAI
  Conference on Artificial Intelligence}, vol.~35, no.~9, 2021, pp. 7987--7994.

\bibitem{chang2020taming}
W.-C. Chang, H.-F. Yu, K.~Zhong, Y.~Yang, and I.~S. Dhillon, ``Taming
  pretrained transformers for extreme multi-label text classification,'' in
  \emph{Proceedings of the 26th ACM SIGKDD international conference on
  knowledge discovery \& data mining}, 2020, pp. 3163--3171.

\bibitem{patra2021semantic}
J.~Patra and M.~Pradel, ``Semantic bug seeding: a learning-based approach for
  creating realistic bugs,'' in \emph{Proceedings of the 29th ACM Joint Meeting
  on European Software Engineering Conference and Symposium on the Foundations
  of Software Engineering}, 2021, pp. 906--918.

\bibitem{tan2014bug}
L.~Tan, C.~Liu, Z.~Li, X.~Wang, Y.~Zhou, and C.~Zhai, ``Bug characteristics in
  open source software,'' \emph{Empirical software engineering}, vol.~19, pp.
  1665--1705, 2014.

\bibitem{liu2019avatar}
K.~Liu, A.~Koyuncu, D.~Kim, and T.~F. Bissyand{\'e}, ``Avatar: Fixing semantic
  bugs with fix patterns of static analysis violations,'' in \emph{2019 IEEE
  26th International Conference on Software Analysis, Evolution and
  Reengineering (SANER)}.\hskip 1em plus 0.5em minus 0.4em\relax IEEE, 2019,
  pp. 1--12.

\bibitem{codeforces}
``Codeforces,'' \url{https://codeforces.com/}, 2023, accessed: 2023-04-01.

\bibitem{lukasczyk2022pynguin}
S.~Lukasczyk and G.~Fraser, ``Pynguin: Automated unit test generation for
  python,'' in \emph{Proceedings of the ACM/IEEE 44th International Conference
  on Software Engineering: Companion Proceedings}, 2022, pp. 168--172.

\bibitem{lahtinen2005study}
E.~Lahtinen, K.~Ala-Mutka, and H.-M. J{\"a}rvinen, ``A study of the
  difficulties of novice programmers,'' \emph{Acm sigcse bulletin}, vol.~37,
  no.~3, pp. 14--18, 2005.

\bibitem{kelleher2005lowering}
C.~Kelleher and R.~Pausch, ``Lowering the barriers to programming: A taxonomy
  of programming environments and languages for novice programmers,'' \emph{ACM
  Computing Surveys (CSUR)}, vol.~37, no.~2, pp. 83--137, 2005.

\bibitem{perretta2022use}
J.~Perretta, A.~DeOrio, A.~Guha, and J.~Bell, ``On the use of mutation analysis
  for evaluating student test suite quality,'' in \emph{Proceedings of the 31st
  ACM SIGSOFT International Symposium on Software Testing and Analysis}, 2022,
  pp. 263--275.

\bibitem{yi2017feasibility}
J.~Yi, U.~Z. Ahmed, A.~Karkare, S.~H. Tan, and A.~Roychoudhury, ``A feasibility
  study of using automated program repair for introductory programming
  assignments,'' in \emph{Proceedings of the 2017 11th Joint Meeting on
  Foundations of Software Engineering}, 2017, pp. 740--751.

\bibitem{bhatia2018neuro}
S.~Bhatia, P.~Kohli, and R.~Singh, ``Neuro-symbolic program corrector for
  introductory programming assignments,'' in \emph{Proceedings of the 40th
  International Conference on Software Engineering}, 2018, pp. 60--70.

\bibitem{hu2019re}
Y.~Hu, U.~Z. Ahmed, S.~Mechtaev, B.~Leong, and A.~Roychoudhury, ``Re-factoring
  based program repair applied to programming assignments,'' in \emph{2019 34th
  IEEE/ACM International Conference on Automated Software Engineering
  (ASE)}.\hskip 1em plus 0.5em minus 0.4em\relax IEEE, 2019, pp. 388--398.

\bibitem{deng2023large}
Y.~Deng, C.~S. Xia, C.~Yang, S.~D. Zhang, S.~Yang, and L.~Zhang, ``Large
  language models are edge-case fuzzers: Testing deep learning libraries via
  fuzzgpt,'' \emph{arXiv preprint arXiv:2304.02014}, 2023.

\bibitem{schafer2023adaptive}
M.~Sch{\"a}fer, S.~Nadi, A.~Eghbali, and F.~Tip, ``Adaptive test generation
  using a large language model,'' \emph{arXiv preprint arXiv:2302.06527}, 2023.

\bibitem{widyasari2020bugsinpy}
R.~Widyasari, S.~Q. Sim, C.~Lok, H.~Qi, J.~Phan, Q.~Tay, C.~Tan, F.~Wee, J.~E.
  Tan, Y.~Yieh \emph{et~al.}, ``Bugsinpy: a database of existing bugs in python
  programs to enable controlled testing and debugging studies,'' in
  \emph{Proceedings of the 28th ACM joint meeting on european software
  engineering conference and symposium on the foundations of software
  engineering}, 2020, pp. 1556--1560.

\bibitem{xia2023keep}
C.~S. Xia and L.~Zhang, ``Keep the conversation going: Fixing 162 out of 337
  bugs for \$0.42 each using chatgpt,'' \emph{arXiv preprint arXiv:2304.00385},
  2023.

\bibitem{liu2023your}
J.~Liu, C.~S. Xia, Y.~Wang, and L.~Zhang, ``Is your code generated by chatgpt
  really correct? rigorous evaluation of large language models for code
  generation,'' \emph{arXiv preprint arXiv:2305.01210}, 2023.

\bibitem{open-coding}
J.~W. Creswell, \emph{{Qualitative Inquiry and Research Design: Choosing Among
  Five Approaches (3rd Edition)}}.\hskip 1em plus 0.5em minus 0.4em\relax SAGE
  Publications, Inc., 2013.

\bibitem{landis1977measurement}
J.~R. Landis and G.~G. Koch, ``The measurement of observer agreement for
  categorical data,'' \emph{biometrics}, pp. 159--174, 1977.

\bibitem{feng2023prompting}
S.~Feng and C.~Chen, ``Prompting is all your need: Automated android bug replay
  with large language models,'' \emph{arXiv preprint arXiv:2306.01987}, 2023.

\bibitem{ebtekar2021elo}
A.~Ebtekar and P.~Liu, ``An elo-like system for massive multiplayer
  competitions,'' \emph{arXiv preprint arXiv:2101.00400}, 2021.

\bibitem{codeforceslevel}
``How to interpret contest ratings,''
  \url{https://codeforces.com/blog/entry/68288}, 2023, accessed: 2023-04-01.

\bibitem{weimer2009automatically}
W.~Weimer, T.~Nguyen, C.~Le~Goues, and S.~Forrest, ``Automatically finding
  patches using genetic programming,'' in \emph{2009 IEEE 31st International
  Conference on Software Engineering}.\hskip 1em plus 0.5em minus 0.4em\relax
  IEEE, 2009, pp. 364--374.

\bibitem{gpt4}
\BIBentryALTinterwordspacing
OpenAI, ``Gpt-4 technical report,'' 2023. [Online]. Available:
  \url{https://arxiv.org/pdf/2303.08774.pdf}
\BIBentrySTDinterwordspacing

\bibitem{fraser2011evosuite}
G.~Fraser and A.~Arcuri, ``Evosuite: automatic test suite generation for
  object-oriented software,'' in \emph{Proceedings of the 19th ACM SIGSOFT
  symposium and the 13th European conference on Foundations of software
  engineering}, 2011, pp. 416--419.

\bibitem{baldoni2018survey}
R.~Baldoni, E.~Coppa, D.~C. D’elia, C.~Demetrescu, and I.~Finocchi, ``A
  survey of symbolic execution techniques,'' \emph{ACM Computing Surveys
  (CSUR)}, vol.~51, no.~3, pp. 1--39, 2018.

\bibitem{23}
W.~Jin and A.~Orso, ``Bugredux: Reproducing field failures for in-house
  debugging,'' in \emph{2012 34th International Conference on Software
  Engineering (ICSE)}.\hskip 1em plus 0.5em minus 0.4em\relax IEEE, 2012, pp.
  474--484.

\bibitem{24}
M.~Soltani, P.~Derakhshanfar, A.~Panichella, X.~Devroey, A.~Zaidman, and A.~van
  Deursen, ``Single-objective versus multi-objectivized optimization for
  evolutionary crash reproduction,'' in \emph{Search-Based Software
  Engineering: 10th International Symposium, SSBSE 2018, Montpellier, France,
  September 8-9, 2018, Proceedings 10}.\hskip 1em plus 0.5em minus 0.4em\relax
  Springer, 2018, pp. 325--340.

\bibitem{25}
M.~Soltani, P.~Derakhshanfar, X.~Devroey, and A.~Van~Deursen, ``A
  benchmark-based evaluation of search-based crash reproduction,''
  \emph{Empirical Software Engineering}, vol.~25, pp. 96--138, 2020.

\bibitem{zhang2022improving}
Z.~Zhang, Y.~Lei, X.~Mao, M.~Yan, and X.~Xia, ``Improving fault localization
  using model-domain synthesized failing test generation,'' in \emph{2022 IEEE
  International Conference on Software Maintenance and Evolution
  (ICSME)}.\hskip 1em plus 0.5em minus 0.4em\relax IEEE, 2022, pp. 199--210.

\bibitem{27}
G.~An and S.~Yoo, ``Human-in-the-loop fault localisation using efficient test
  prioritisation of generated tests,'' \emph{arXiv preprint arXiv:2104.06641},
  2021.

\bibitem{manes2019art}
V.~J. Man{\`e}s, H.~Han, C.~Han, S.~K. Cha, M.~Egele, E.~J. Schwartz, and
  M.~Woo, ``The art, science, and engineering of fuzzing: A survey,''
  \emph{IEEE Transactions on Software Engineering}, vol.~47, no.~11, pp.
  2312--2331, 2019.

\bibitem{saha2023rare}
S.~Saha, L.~Sarker, M.~Shafiuzzaman, C.~Shou, A.~Li, G.~Sankaran, and
  T.~Bultan, ``Rare path guided fuzzing,'' in \emph{Proceedings of the 32nd ACM
  SIGSOFT International Symposium on Software Testing and Analysis}, 2023, pp.
  1295--1306.

\bibitem{liang2019deepfuzzer}
J.~Liang, Y.~Jiang, M.~Wang, X.~Jiao, Y.~Chen, H.~Song, and K.-K.~R. Choo,
  ``Deepfuzzer: Accelerated deep greybox fuzzing,'' \emph{IEEE Transactions on
  Dependable and Secure Computing}, vol.~18, no.~6, pp. 2675--2688, 2019.

\bibitem{chen2020metamorphic}
T.~Y. Chen, S.~C. Cheung, and S.~M. Yiu, ``Metamorphic testing: a new approach
  for generating next test cases,'' \emph{arXiv preprint arXiv:2002.12543},
  2020.

\bibitem{tian2021extent}
Y.~Tian, S.~Ma, M.~Wen, Y.~Liu, S.-C. Cheung, and X.~Zhang, ``To what extent do
  dnn-based image classification models make unreliable inferences?''
  \emph{Empirical Software Engineering}, vol.~26, no.~5, p.~84, 2021.

\bibitem{cao2022semmt}
J.~Cao, M.~Li, Y.~Li, M.~Wen, S.-C. Cheung, and H.~Chen, ``Semmt: a
  semantic-based testing approach for machine translation systems,'' \emph{ACM
  Transactions on Software Engineering and Methodology (TOSEM)}, vol.~31,
  no.~2, pp. 1--36, 2022.

\bibitem{segura2016survey}
S.~Segura, G.~Fraser, A.~B. Sanchez, and A.~Ruiz-Cort{\'e}s, ``A survey on
  metamorphic testing,'' \emph{IEEE Transactions on software engineering},
  vol.~42, no.~9, pp. 805--824, 2016.

\bibitem{mckeeman1998differential}
W.~M. McKeeman, ``Differential testing for software,'' \emph{Digital Technical
  Journal}, vol.~10, no.~1, pp. 100--107, 1998.

\bibitem{evans2007differential}
R.~B. Evans and A.~Savoia, ``Differential testing: a new approach to change
  detection,'' in \emph{The 6th Joint Meeting on European software engineering
  conference and the ACM SIGSOFT Symposium on the Foundations of Software
  Engineering: Companion Papers}, 2007, pp. 549--552.

\bibitem{chen2016coverage}
Y.~Chen, T.~Su, C.~Sun, Z.~Su, and J.~Zhao, ``Coverage-directed differential
  testing of jvm implementations,'' in \emph{proceedings of the 37th ACM
  SIGPLAN Conference on Programming Language Design and Implementation}, 2016,
  pp. 85--99.

\bibitem{tian2023chatgpt}
H.~Tian, W.~Lu, T.~O. Li, X.~Tang, S.-C. Cheung, J.~Klein, and T.~F.
  Bissyand{\'e}, ``Is chatgpt the ultimate programming assistant--how far is
  it?'' \emph{arXiv preprint arXiv:2304.11938}, 2023.

\bibitem{vaithilingam2022expectation}
P.~Vaithilingam, T.~Zhang, and E.~L. Glassman, ``Expectation vs. experience:
  Evaluating the usability of code generation tools powered by large language
  models,'' in \emph{Chi conference on human factors in computing systems
  extended abstracts}, 2022, pp. 1--7.

\bibitem{ahmed2022few}
T.~Ahmed and P.~Devanbu, ``Few-shot training llms for project-specific
  code-summarization,'' \emph{arXiv preprint arXiv:2207.04237}, 2022.

\bibitem{jalil2023chatgpt}
S.~Jalil, S.~Rafi, T.~D. LaToza, K.~Moran, and W.~Lam, ``Chatgpt and software
  testing education: Promises \& perils,'' \emph{arXiv preprint
  arXiv:2302.03287}, 2023.

\bibitem{pearce2021can}
H.~Pearce, B.~Tan, B.~Ahmad, R.~Karri, and B.~Dolan-Gavitt, ``Can openai codex
  and other large language models help us fix security bugs?'' \emph{arXiv
  preprint arXiv:2112.02125}, 2021.

\bibitem{xia2022practical}
C.~S. Xia, Y.~Wei, and L.~Zhang, ``Practical program repair in the era of large
  pre-trained language models,'' \emph{arXiv preprint arXiv:2210.14179}, 2022.

\end{thebibliography}

\end{document}